\documentclass[reprint,twocolumn]{revtex4-2}
\usepackage{amsfonts}
\usepackage{amsmath}
\usepackage{amssymb}
\usepackage{charter}
\usepackage{graphicx}
\usepackage{float}
\usepackage{amsmath}
\usepackage{amssymb}
\usepackage{charter}
\usepackage{graphicx}
\usepackage{subfigure}
\usepackage{graphicx}
\usepackage{epstopdf}
\usepackage{color}
\usepackage{tocvsec2}
\usepackage{enumerate}
\usepackage{graphicx}
\usepackage{dcolumn}
\usepackage{bm}

\begin{document}

\title{Two-parameter estimation via photon subtraction operation within a
feedback-assisted interferometer }
\author{Qingqian Kang$^{1,2}$}
\author{Zekun Zhao$^{1}$}
\author{Qisi Zhou$^{1}$}
\author{Teng Zhao$^{1}$}
\author{Cunjin Liu$^{1}$}
\author{Xin Su$^{1}$}
\author{Liyun Hu$^{1,*}$}
\author{Sanqiu Liu$^{1,3,}$}
\thanks{Corresponding authors: hlyun@jxnu.edu.cn, sqliu@ncu.edu.cn}
\affiliation{$^{{\small 1}}$\textit{Center for Quantum Science and Technology, Jiangxi
Normal University, Nanchang 330022, China}\\
$^{{\small 2}}$\textit{Department of Physics, College of Science and Technology,
Jiangxi Normal University, Nanchang 330022, China}\\
$^{{\small 3}}$\textit{Department of Physics, Nanchang University, Nanchang 330031, China}}

\begin{abstract}
In this paper, we analyze how multi-photon subtraction operations in a
feedback-assisted interferometer can enhance measurement precision for
single-parameter and two-parameter estimation under both ideal and
photon-loss conditions. We examine the effects of the feedback strength $R$,
the optical parametric amplifier's gain $g$, the coherent state amplitude $%
\alpha $, and the order of multi-photon subtraction on system performance.
We demonstrate that an optimal feedback strength $R_{opt}$ exists in both
conditions. Selecting a suitable $R$ can significantly boost the system's
robustness to photon loss, and markedly improve measurement precision. And
the photon subtraction operations within a feedback-assisted interferometer
can further enhance measurement precision effectively. Additionally, we find
that increasing intramode correlations while decreasing intermode
correlations can improve estimation accuracy. This work investigates a new
method through the synergistic integration of feedback topology and
non-Gaussian operations into a multiparameter estimation system, along with
their systematic study under both ideal and loss conditions. The findings
may contribute to improving quantum-enhanced measurements and hold promise
for high-precision quantum sensing research.

\end{abstract}

\maketitle

\section{Introduction}

Quantum precision measurement technology enjoys a wide range of uses \cite%
{1,2,3,4,5,6,7}. It covers high-accuracy optical frequency standards and
time-frequency transfer, and also applies to quantum gyroscopes, atomic
gravimeters and other quantum navigation technology, as well as quantum
radar, trace atom tracking and weak magnetic field detection in quantum
sensitive detection technology. Phase estimation is a crucial method for
precision measurement that allows the estimation of many physical quantities
not directly measurable by traditional approaches, and accordingly, there
has been extensive research and significant progress in the field of optical
interference measurement.

So far, theoretical and experimental studies have largely focused on
single-phase estimation. Some proof-of-principle experiments have
demonstrated the ability to estimate phases beyond the standard
quantum-noise limit (SQL), including in magnetometry \cite{8}, atomic clocks
\cite{9}, and optical detection of gravitational waves \cite{10}. However,
in practical scenarios such as quantum sensing and imaging \cite{11}, and
quantum communication and computing protocols \cite{12}, the estimation
process often requires the simultaneous measurement of multiple parameters.
This consideration has sparked interest in the theoretical \cite{13,14,15,16}
and experimental \cite{17,18} study of multiparameter quantum estimation.

In recent years, multiparameter estimation (i.e., estimating multiple
parameters simultaneously) has drawn increasing attention. Theoretical
research has demonstrated that simultaneous estimation of multiple phases
holds an advantage over individual optimal phase estimation \cite{18a}. And
this advantage remains even in the presence of photon loss \cite{18b}. Pezz%
\`{e} \textit{et al.} \cite{19} tackled one of the key difficulties of
multiphase estimation: obtaining a measurement which saturates the
fundamental sensitivity bounds. They derived necessary and sufficient
conditions for projective measurements acting on pure states to saturate the
ultimate theoretical bound on precision given by the quantum Fisher
information matrix (QFIM). Nichols \textit{et al.} \cite{20} derived general
analytical expressions for the QFIM and measurement compatibility
conditions. These ensure the asymptotic attainability of the quantum Cram%
\'{e}r-Rao bound (QCRB) when estimating multiple parameters encoded in
multimode Gaussian states. Yang \textit{et al.} \cite{21} extended the
method of Braunstein and Caves to general quantum-state multiparameter
estimation. They derived the matrix bound of the classical Fisher
information matrix for each measurement operator. Ho \textit{et al.} \cite%
{22} presented a framework for simultaneous multiparameter estimation using
quantum sensors in a specific noisy environment. Hou \textit{et al.} \cite%
{23} experimentally demonstrated an optimally controlled multipass scheme
that saturates three Heisenberg uncertainty relations, achieving maximum
precision for the estimation of all three parameters in SU(2) operators. The
experiment achieved ultimate precision for multiple parameters
simultaneously, marking a key advance in multiparameter quantum metrology.

However, translating theoretical advantages into practical performance faces
a crucial challenge: the vulnerability of quantum resources in measurement
systems to environmental dissipation. Taking the representative SU(1,1)
nonlinear interferometer as an example, while it demonstrates potential for
surpassing the SQL in single-parameter measurements, internal photon losses
significantly degrade quantum nonclassicality, creating bottlenecks for
precision enhancement in multiparameter scenarios. To address this,
researchers proposed two pathways: The first involves quantum state
engineering through non-Gaussian operations \cite{24,25,26,27}. Photon
subtraction (PS), photon addition (PA), photon catalysis (PC), quantum
scissor (QS), and their coherent superposition effectively increase the
non-classicality and entanglement degrees of quantum states. This, in turn,
enhances their potential in quantum information processing \cite{28}. The
second pathway focuses on topological innovation in interferometer
architectures. Taking the feedback optical parametric amplifier (FOPA) \cite%
{h1} as an example, it has been experimentally realized by using the
four-wave mixing (FWM) process as the underlying optical parametric
amplifier (OPA) and the beam splitter (BS) as the feedback controller \cite%
{29}. The FOPA demonstrates unique quantum correlation and\ entanglement
enhancement through feedback mechanisms compared to the traditional optical
parametric amplifier (TOPA) \cite{30,h3}.

Notably, existing research has not yet to fully exploit the synergistic
potential of these two strategies. Although existing studies have
demonstrated the unique advantages of FOPA over the TOPA, its applications
remain confined to improving single-parameter estimation under ideal
condition \cite{h2}. Meanwhile, while research on non-Gaussian operations
has explored various sophisticated schemes, there has been no analysis of
their integration with feedback-assisted interferometers for joint
multiparameter optimization. This research gap urgently needs to be
addressed due to the following critical insights: Multiparameter estimation
demands systems to simultaneously sustain high-dimensional quantum
correlations and dissipation-resistant robustness. Crucially, the feedback
mechanism of FOPA enables active control over parameter coupling effects,
while non-Gaussian operations can strategically compensate for loss-induced
decoherence. The synergistic integration of these two approaches holds the
potential to surpass existing precision limits by leveraging their
complementary advantages in parameter controllability and quantum resource
protection.

Building on these insights, this work pioneers the synergistic incorporation
of FOPA topology and multi-PS operations into multiphase estimation systems,
systematically investigating their performance under both ideal and lossy
conditions. By constructing analytical QFIM for two-parameter estimation, we
reveal that this hybrid scheme can significantly improve the phase
measurement accuracy of two-parameter. Compared to conventional schemes, our
scheme demonstrates outstanding robustness against photon losses in both
single-parameter and two-parameter estimation scenarios. These results
provide a new paradigm for developing practical multiparameter quantum
sensors.

The paper is structured as follows: Sec. II develops the theoretical model
of the FOPA. Sec. III concentrates on single-parameter estimation of
multi-PS operations in a feedback-assisted interferometer, analyzing the
improvement in the QCRB under both ideal and loss conditions. Sec. IV
broadens the perspective to two-parameter estimation, exploring the QCRB
improvement under the same conditions. Sec. V emphasizes the system's
intramode correlation and intermode correlation characteristics. Lastly,
Sec. VI summarizes the entire paper.

\section{The model of optical parametric amplifier with feedback}

\bigskip The topology of FOPA is shown in Fig. 1. The FOPA is composed of
one OPA and two BSs (BS$_{1}$ and BS$_{2}$). Here, the OPA functions as the
entanglement resource, while the BSs act as feedback controllers. Injecting
a pump mode ($P$) into the OPA creates an entangled state \{$\hat{a}_{2} $, $%
\hat{b}_{2} $\}.

\begin{figure}[tph]
\label{Fig1} \centering \includegraphics[width=0.9\columnwidth]{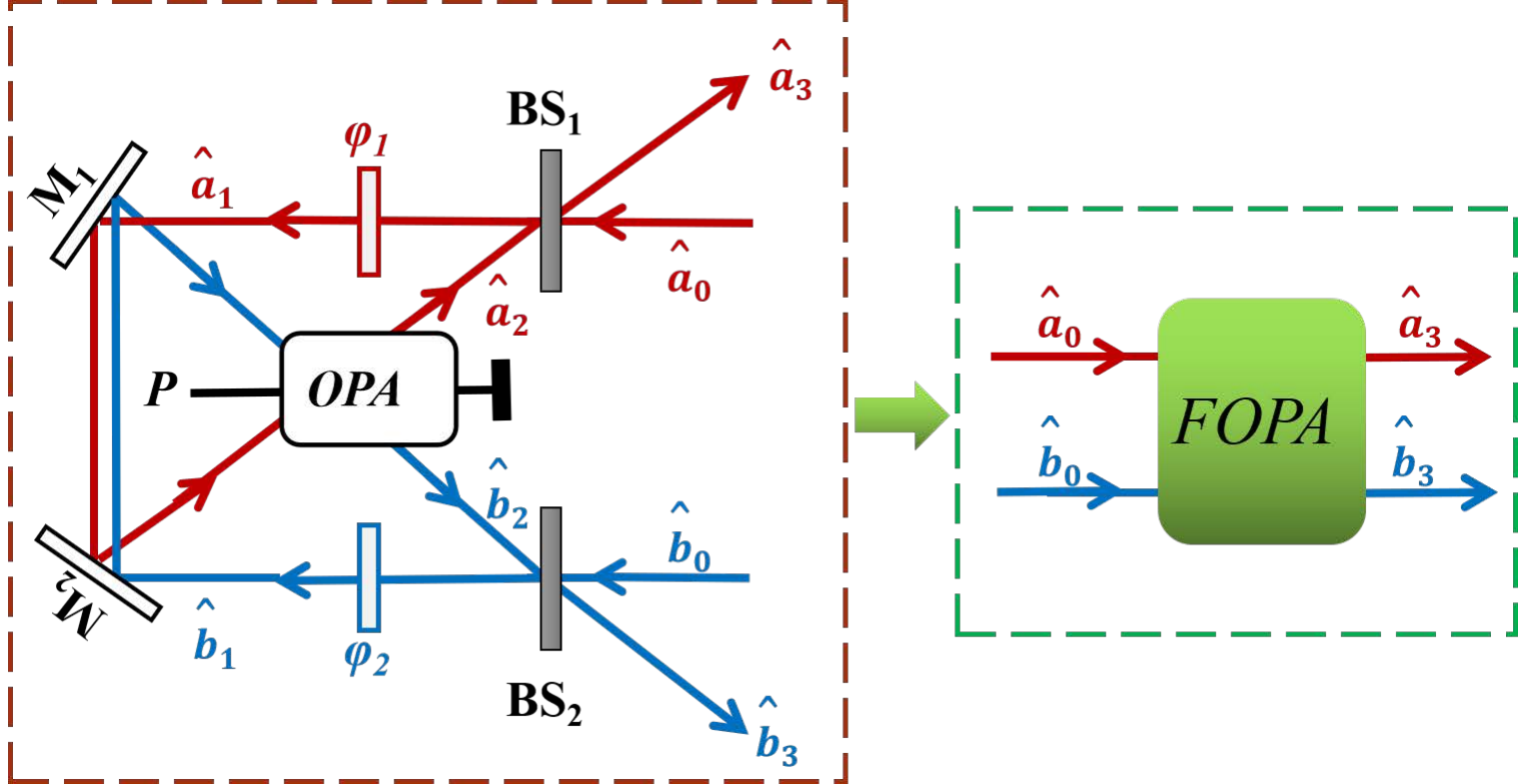}
\caption{Schematic diagram of the FOPA. It consists of an OPA and two BSs (BS%
$_{1}$ and BS$_{2}$), with the two output states of the OPA being coherently
feedback into the two input ports via BS$_{1}$ and BS$_{2}$, respectively.
BS, beam splitter; M, mirror.}
\end{figure}

The input-output relationship of OPA is given by \cite{h1,h2}
\begin{eqnarray}
\hat{a}_{2} &=&G\hat{a}_{1}e^{i\varphi _{1}}+g\hat{b}_{1}^{\dagger
}e^{-i\varphi _{2}},  \notag \\
\hat{b}_{2} &=&G\hat{b}_{1}e^{i\varphi _{2}}+g\hat{a}_{1}^{\dagger
}e^{-i\varphi _{1}},  \label{1}
\end{eqnarray}%
where $g$ is the gain factor of the OPA with $G^{2}-g^{2}=1$. $\varphi _{1}$
and $\varphi _{2}$ denote the phase shifts introduced by the feedback paths.
After generating $\hat{a}_{2}$ and $\hat{b}_{2}$, they are sent to the BSs.
The input-output relationships of BS$_{1}$ and BS$_{2}$ are given by%
\begin{eqnarray}
\hat{a}_{1} &=&\sqrt{1-R_{1}}\hat{a}_{0}-\sqrt{R_{1}}\hat{a}_{2},  \notag \\
\hat{a}_{3} &=&\sqrt{R_{1}}\hat{a}_{0}+\sqrt{1-R_{1}}\hat{a}_{2},  \label{2}
\end{eqnarray}%
and%
\begin{eqnarray}
\hat{b}_{1} &=&\sqrt{1-R_{2}}\hat{b}_{0}-\sqrt{R_{2}}\hat{b}_{2},  \notag \\
\hat{b}_{3} &=&\sqrt{R_{2}}\hat{b}_{0}+\sqrt{1-R_{2}}\hat{b}_{2}.  \label{3}
\end{eqnarray}%
In these above equations, $R_{i}$ represents the reflectivity of BS$_{i}$ ($%
i=1,2$). The BS$_{1}$ and BS$_{2}$ act as controllers, coherently feeding $%
\hat{a}_{2}$ and $\hat{b}_{2}$ back into their input ports and thus
affecting the input-output relationship of the OPA. The feedback strength is
determined by the reflectivity $R_{i}$. Consequently, the input-output
relation of the FOPA is described by \cite{h3}
\begin{equation}
\hat{a}_{3}=\frac{k_{1}}{k_{0}}\hat{a}_{0}+\frac{k_{3}}{k_{0}}\hat{b}%
_{0}^{\dagger },b_{3}^{\dagger }=\frac{k_{2}}{k_{0}}\hat{b}_{0}^{\dagger }+%
\frac{k_{4}}{k_{0}}\hat{a}_{0},  \label{4}
\end{equation}%
with%
\begin{eqnarray}
k_{0} &=&1+G\left( \sqrt{R_{1}}e^{i\varphi _{1}}+\sqrt{R_{2}}e^{-i\varphi
_{2}}\right) +\sqrt{R_{1}R_{2}}e^{i\left( \varphi _{1}-\varphi _{2}\right) },
\notag \\
k_{1} &=&\sqrt{R_{1}}+G\left( \sqrt{R_{1}R_{2}}e^{-i\varphi
_{2}}+e^{i\varphi _{1}}\right) +\sqrt{R_{2}}e^{i\left( \varphi _{1}-\varphi
_{2}\right) },  \notag \\
k_{2} &=&\sqrt{R_{2}}+G\left( \sqrt{R_{1}R_{2}}e^{i\varphi
_{1}}+e^{-i\varphi _{2}}\right) +\sqrt{R_{1}}e^{i\left( \varphi _{1}-\varphi
_{2}\right) },  \notag \\
k_{3} &=&g\sqrt{\left( 1-R_{1}\right) \left( 1-R_{2}\right) }e^{-i\varphi
_{2}},  \notag \\
k_{4} &=&g\sqrt{\left( 1-R_{1}\right) \left( 1-R_{2}\right) }e^{i\varphi
_{1}}.  \label{5}
\end{eqnarray}%
When $R_{i}=0$, the FOPA is reduced to the TOPA. We use the signal flow
diagram method and apply Mason's gain formula to calculate the input-output
relationship of FOPA, and the detailed derivation is in Appendix A.

Previous studies have demonstrated that in a FOPA system, setting the phase
delays $\varphi _{1}$ and $\varphi _{2}$ of the feedback path to $\pi $ can
achieve optimal entanglement enhancement \cite{h3}. In quantum optics, when
the phase is $\pi $, the feedback signal is out of phase with the input
signal. This negative feedback mechanism helps to preserve the strength of
quantum entanglement and enhances the precision of system control. Even in
the presence of losses, negative feedback can improve the system's
robustness by mitigating some of the noise caused by the losses. Therefore,
setting $\varphi _{1}$ and $\varphi _{2}$ to $\pi $ in the feedback path can
optimize the feedback control and consequently enhance the overall
performance of the system. For simplicity, we set $R_{1}=R_{2}=R$ and $%
\varphi _{1}=\varphi _{2}=\pi $.

\section{Single-parameter estimation}

\bigskip The model of the single-parameter estimation with FOPA is shown in
Fig. 2. The input states are coherent state $\left \vert \alpha
\right
\rangle _{a}$ and vacuum state $\left \vert 0\right \rangle _{b}$.
Based on Eqs. (\ref{4}) and (\ref{5}), we can obtain the transformation
relationship of the FOPA, that is,
\begin{eqnarray}
\hat{U}_{F}^{\dagger }\hat{a}\hat{U}_{F} &=&\frac{k_{1}}{k_{0}}\hat{a}+\frac{%
k_{3}}{k_{0}}\hat{b}^{\dagger },  \notag \\
\hat{U}_{F}^{\dagger }\hat{a}^{\dagger }\hat{U}_{F} &=&\frac{k_{1}^{\ast }}{%
k_{0}^{\ast }}\hat{a}^{\dagger }+\frac{k_{3}^{\ast }}{k_{0}^{\ast }}\hat{b},
\notag \\
\hat{U}_{F}^{\dagger }\hat{b}\hat{U}_{F} &=&\frac{k_{2}^{\ast }}{k_{0}^{\ast
}}\hat{b}+\frac{k_{4}^{\ast }}{k_{0}^{\ast }}\hat{a}^{\dagger },  \notag \\
\hat{U}_{F}^{\dagger }\hat{b}^{\dagger }\hat{U}_{F} &=&\frac{k_{2}}{k_{0}}%
\hat{b}^{\dagger }+\frac{k_{4}}{k_{0}}\hat{a},  \label{6}
\end{eqnarray}%
where, $\hat{U}_{F}$ represents the operator of the FOPA. After the FOPA,
mode $a$ experiences a phase shift process $\hat{U}_{\phi }=\exp [-i\phi
_{a}(\hat{a}^{\dagger }\hat{a})]$, while mode $b$ remains unaltered. The
transformation relationship of the phase shifter $\hat{U}_{\phi }$ is as
follows%
\begin{equation}
\hat{U}_{\phi }^{\dagger }\hat{a}\hat{U}_{\phi }=\hat{a}e^{i\phi _{a}}.
\label{7}
\end{equation}%
The operators of two fictitious BSs can be represented as $\hat{U}_{B_{1}}\ $%
and $\hat{U}_{B_{2}}$, with $\hat{U}_{B_{1}}=\exp \left[ \theta
_{T_{1}}\left( \hat{a}^{\dagger }\hat{a}_{v_{1}}-\hat{a}\hat{a}%
_{v_{1}}^{\dagger }\right) \right] $ and $\hat{U}_{B_{2}}=\exp \left[ \theta
_{T_{2}}\left( \hat{a}^{\dagger }\hat{a}_{v_{2}}-\hat{a}\hat{a}%
_{v_{2}}^{\dagger }\right) \right] $, where $\hat{a}_{v_{1}}$ and $\hat{a}%
_{v_{2}}$ represent vacuum modes. Here, $T_{k}$ ($k=1,2$) denotes the
transmissivity of the fictitious BSs, which is associated with $\theta
_{T_{k}}$ through $T_{k}=\cos ^{2}\theta _{T_{k}}\in \left[ 0,1\right] $.
The value of transmittance equal to $1$ ($T_{k}=1$) corresponds to the ideal
case without photon losses \cite{h4}. To improve the measurement accuracy,
we introduced a multi-PS operation after the FOPA. In the multi-PS, $m$ and $%
n$ photons are subtracted from mode $a$ and mode $b$, respectively. This
process can be described by the operator $\hat{U}_{P}=\hat{a}^{m}\otimes
\hat{b}^{n}$.

\begin{figure}[tph]
\label{Fig2} \centering \includegraphics[width=0.9\columnwidth]{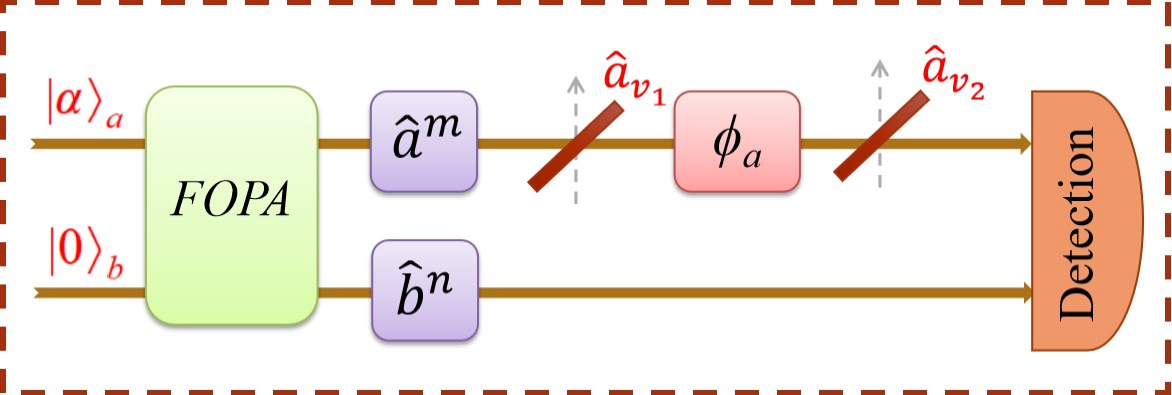}
\caption{Schematic diagram of the single-parameter estimation. The two input
ports are a coherent state $\left \vert \protect \alpha \right \rangle _{a}$
and a vacuum state $\left \vert 0\right \rangle _{b}$. $\hat{a}_{v_{1}}$ and
$\hat{a}_{v_{2}}$ are vacuum modes. $\protect \phi _{a}$ is the phase
shifter. }
\end{figure}

\subsection{\protect \bigskip Ideal case}

\bigskip Initially, we consider the ideal case, $T_{k}=1$ (where $k=1,2$),
representing the case without photon losses. For a pure state system, the
QFI can be derived by \cite{h5}%
\begin{equation}
F_{a}=4\left[ \left \langle \Psi ^{\prime }|\Psi ^{\prime }\right \rangle
-\left \vert \left \langle \Psi ^{\prime }|\Psi \right \rangle \right \vert
^{2}\right] ,  \label{8}
\end{equation}%
where\ $\left \vert \Psi \right \rangle $ is the quantum state after the
phase shift and before the detection, and $\left \vert \Psi ^{\prime
}\right
\rangle =\partial \left \vert \Psi \right \rangle /\partial \phi
_{a}.$ Then the QFI can be reformulated as \cite{h5}
\begin{equation}
F_{a}=4\left \langle \Delta ^{2}\hat{n}_{a}\right \rangle ,  \label{9}
\end{equation}%
where $\hat{n}_{a}=\hat{a}^{\dag }\hat{a}$, $\left \langle \Delta ^{2}\hat{n}%
_{a}\right \rangle =\left \langle \Psi \right \vert (\hat{a}^{\dagger }\hat{a%
})^{2}|\Psi \rangle -\left( \left \langle \Psi \right \vert \hat{a}^{\dagger
}\hat{a}|\Psi \rangle \right) ^{2}$.

In the ideal case, the quantum state is expressed as $\left \vert \Psi
\right \rangle =A\hat{U}_{\phi }\hat{U}_{P}\hat{U}_{F}\left \vert \alpha
\right \rangle \left \vert 0\right \rangle $. Thus, the QFI can be expressed
as
\begin{eqnarray}
F_{a} &=&4[A^{2}\Gamma _{m,n,2,2,0,0}+A^{2}\Gamma _{m,n,1,1,0,0}  \notag \\
&&-\left( A^{2}\Gamma _{m,n,1,1,0,0}\right) ^{2}],  \label{10}
\end{eqnarray}%
where the expression for $\Gamma _{^{m,n,x_{1},y_{1},x_{2},y_{2}}}$ is given
in Appendix B. The normalization constant $A$ for the multi-PS, is given by
\cite{h4}
\begin{equation}
A^{-2}=\Gamma _{^{m,n,0,0,0,0}}.  \label{11}
\end{equation}%
We can explore the connection between the QFI and the related parameters
through the use of Eq. (\ref{10}).

The QCRB, as defined by Ref. \cite{h6}, represents the minimum phase
sensitivity achievable across all measurement schemes:%
\begin{equation}
QCRB_{a}=\frac{1}{\sqrt{vF_{a}}},  \label{12}
\end{equation}%
where, $v$ denotes the number of measurements. For simplicity, we set $v=1$.
The QCRB \cite{h6,h7} establishes the ultimate limit for probabilities
derived from quantum system measurements. It serves as an estimator
implemented asymptotically via maximum likelihood estimation and provides a
measurement-independent phase sensitivity.

To evaluate the optimality of the phase sensitivity achieved by the FOPA
with multi-PS operations, we analyze the variation curve of $QCRB_{a}$ with
the feedback coefficient $R$ in Fig. 3. The red solid line indicates the
sensitivity without multi-PS and feedback, corresponding to $R=0$ and $m=n=0$%
. Compared to no feedback, the phase sensitivity without multi-PS ($m=n=0$)
is improved when $R<0.5$. However, as the order of the multi-PS operations
increases, not only can the measurement accuracy be further improved, but
also the range of advantage for $R$ can be expanded. This shows that
non-Gaussian operations can further enhance the precision of phase
measurement under the new model structure.

Without non-Gaussian operations, when $\varphi _{1}=\varphi _{2}=\pi $, the
optimal reflectivity can be calculated using the formula from Refs. \cite%
{h2,h3}, i.e.,
\begin{equation}
R_{opt}=1+2g^{2}-2g\sqrt{1+g^{2}}.  \label{13}
\end{equation}%
As shown in Figs. 3 and 4, the value of QCRB initially decreases and then
increases with $R$, indicating that there is an optimal $R$ for the best
interferometer performance. Numerical simulations confirm that even with
non-Gaussian operations, the system maintains the same $R_{opt}\sim g$
relationship as in Eq. (\ref{13}). The optimal feedback strength $R_{opt}$
varies with different gain factor $g$, consistent with Refs. \cite{h2,h3}.
This is attributed to the maximal entanglement of the FOPA leading to the
best phase sensitivity \cite{h2}.

When $g=1$, substituting this condition into Eq. (\ref{13}) yields an
optimal reflectivity $R_{opt}=3-2\sqrt{2}$, which is approximately $0.17$.
This finding is visually represented in Fig. 3. Notably, upon closer
examination of the graph, it is observed that when the reflectivity $R$
attains the optimal value $R_{opt}$, this point exhibits a singularity.
Further analysis reveals that this peculiar phenomenon arises due to the
denominator in the coefficients becoming zero when calculating the
input-output relationship of the FOPA at the optimal reflectivity $R_{opt}$.
This suggests that under specific conditions, the system may enter a
critical state, which holds significant implications for comprehending the
system's physical properties and enhancing its performance.

\begin{figure}[tph]
\label{Fig3} \centering \includegraphics[width=0.9\columnwidth]{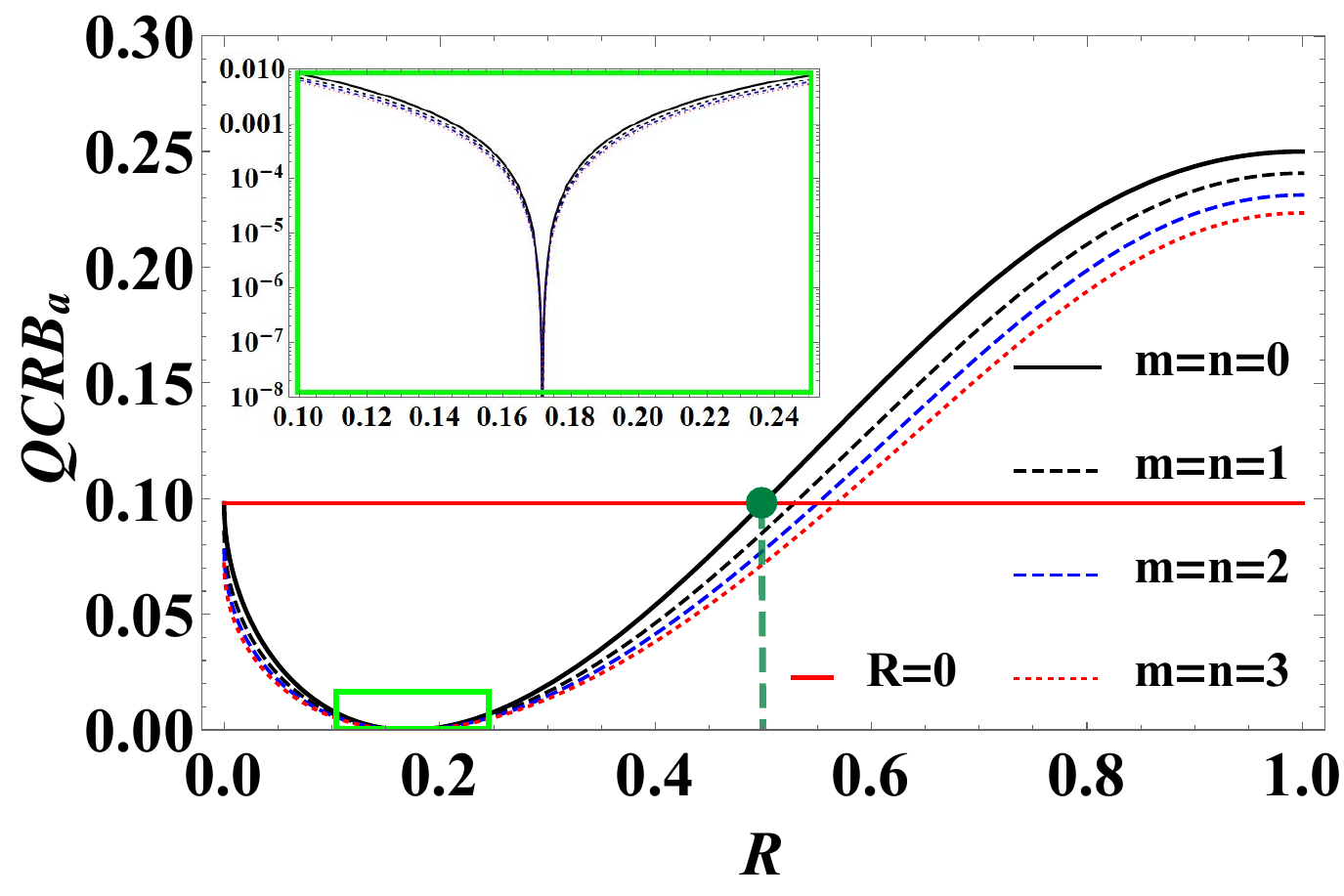}
\caption{The $QCRB_{a}$ as a function of $R$, with $g=1$ and $\protect \alpha %
=2$. The red solid line represents the sensitivity of the TOPA without
multi-PS operation, serving as a reference. $m$ and $n$ denote the order of
the multi-PS operations from mode $a$ and mode $b$, respectively.}
\end{figure}

To verify the singular point's system impact, we further plot the total
average photon number $N$ versus feedback strength $R$ in Fig. 4. From the
figure, we can draw the following conclusions: without multi-PS operations,
the FOPA can increase $N$ only when $R<0.5$. Moreover, as the order of
multi-PS increases, the photon number increase becomes larger, and the range
of $R$'s advantage is expanded. Similarly, a singular point occurs around $%
R\approx 0.17$, where the total average photon number $N$ suddenly increases
sharply and tends to infinity when $R$ approaches the optimal value $R_{opt}$%
.

\begin{figure}[tph]
\label{Fig4} \centering \includegraphics[width=0.9\columnwidth]{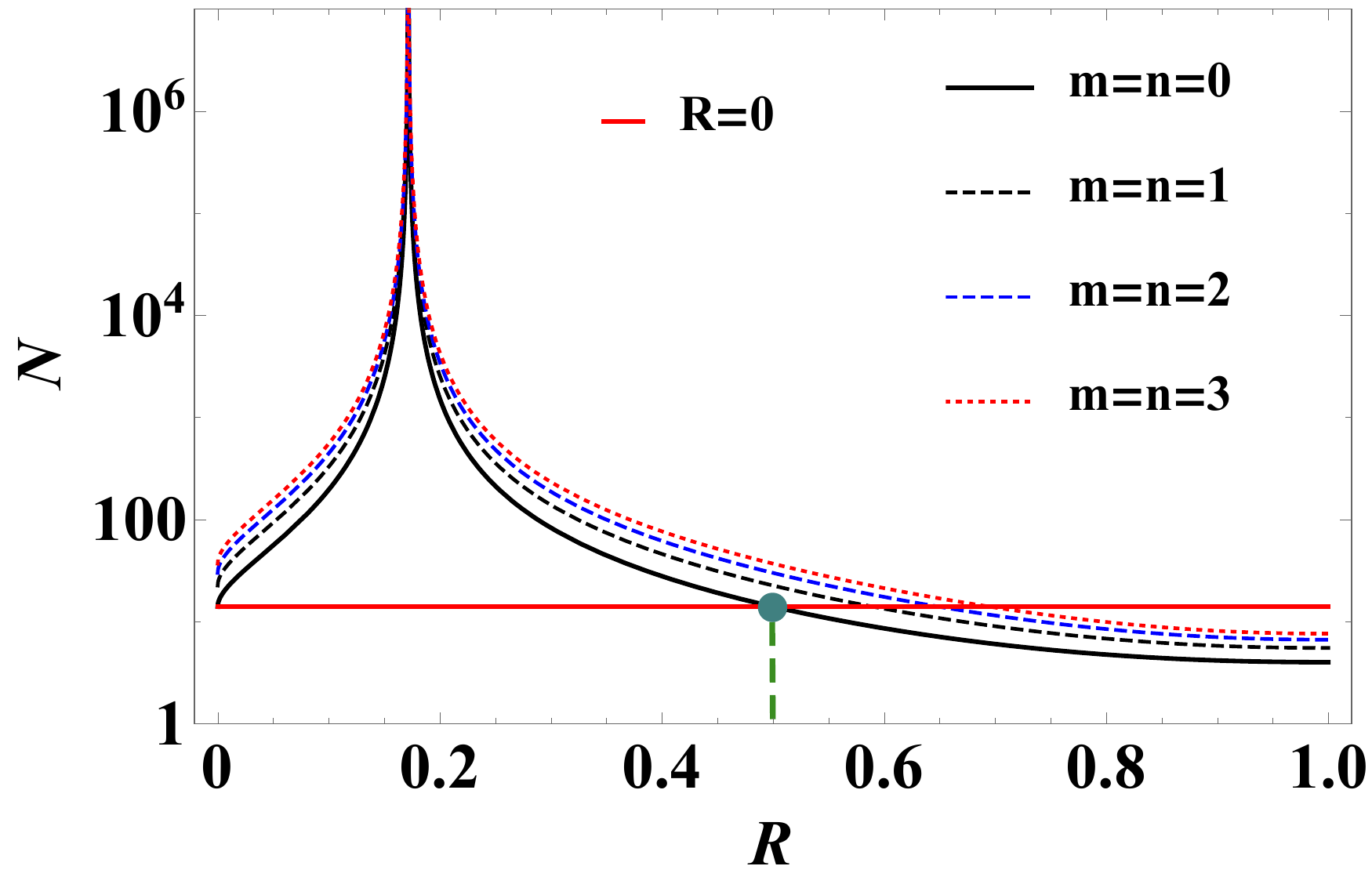}
\caption{The $N$ as a function of $R$, with $g=1$ and $\protect \alpha =2$.
The red solid line represents the sensitivity of the TOPA without multi-PS
operation, serving as a reference. $m$ and $n$ denote the order of the
multi-PS operations from mode $a$ and mode $b$, respectively.}
\end{figure}

To investigate the model parameters' effects on the system, we plot the $%
QCRB_{a}$ curves against $g$ and $\alpha $ as shown in Figs. 5 and 6. It is
shown that $QCRB_{a}$ improves with increasing $g$ and $\alpha $. The
feedback system demonstrates a significantly higher sensitivity compared to
the system without feedback. Moreover, as the order of multi-PS increases,
further improvement in $QCRB_{a}$ is achieved.

\begin{figure}[tph]
\label{Fig5} \centering \includegraphics[width=0.9\columnwidth]{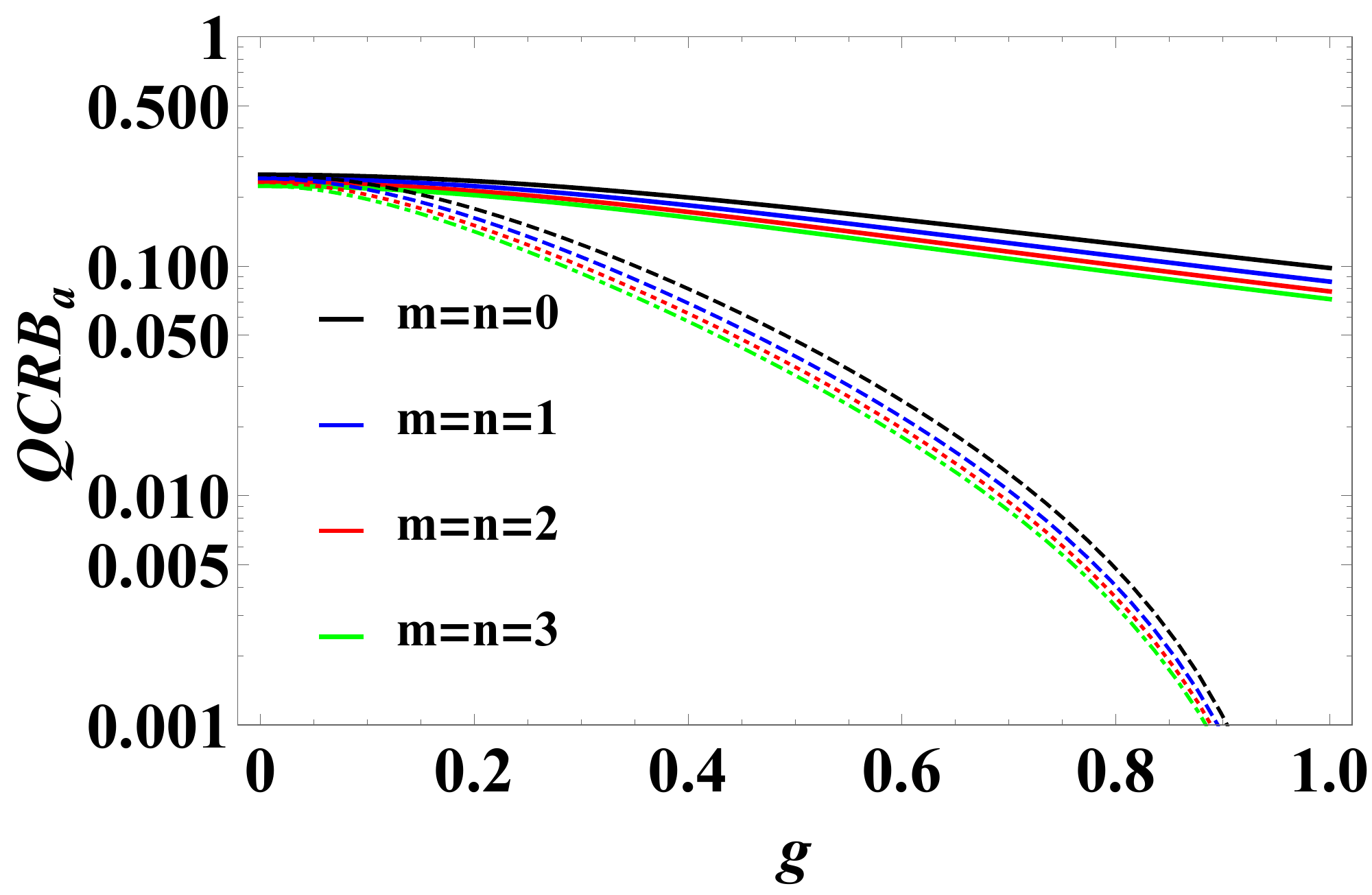}
\caption{The $QCRB_{a}$ as a function of $g$, with $R=0.17$ and $\protect%
\alpha =2$. The solid line corresponds to the system without feedback (i.e.,
$R=0$); The dashed line corresponds to the FOPA system ($R=0.17$). $m$ and $%
n $ denote the order of the multi-PS operations from mode $a$ and mode $b$,
respectively.}
\end{figure}

\begin{figure}[tph]
\label{Fig6} \centering \includegraphics[width=0.9\columnwidth]{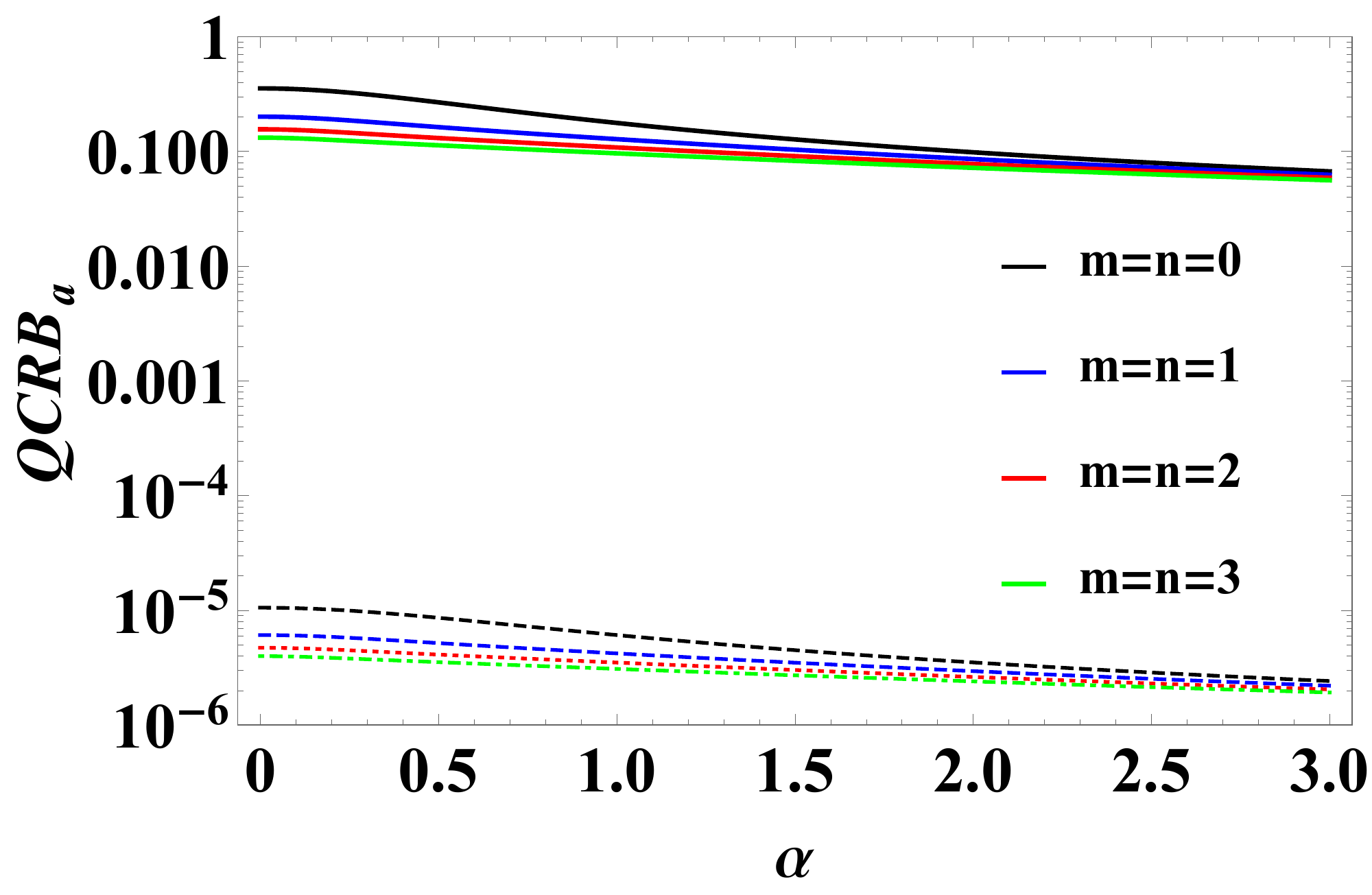}
\caption{The $QCRB_{a}$ as a function of $\protect \alpha $, with $R=0.17$
and $g=1$. The solid line corresponds to the system without feedback (i.e., $%
R=0$); The dashed line corresponds to the FOPA system ($R=0.17$). $m$ and $n$
denote the order of the multi-PS operations from mode $a$ and mode $b$,
respectively.}
\end{figure}

\subsection{\protect \bigskip Photon-loss case}

\bigskip In phase measurement, precision is highly susceptible to photon
loss, particularly internal loss. Thus, we extend our analysis to include
the QCRB in the presence of photon loss, corresponding to $T_{k}\in (0,1)$,
as shown in Fig. 2. For realistic quantum systems, Escher \textit{et al}.
\cite{h5} proposed a method for calculating the QFI. This method can be
briefly summarized as follows.

The QFI with photon loss is calculated as detailed in Ref. \cite{h5}. After
the FOPA $\hat{U}_{F}$, multi-PS $\hat{U}_{P}$, phase shift $\hat{U}_{\phi }$%
, photon loss $\hat{U}_{B}$, and before detection, the output state in an
expanded space is given by
\begin{equation}
\left \vert \Psi _{S}\right \rangle =\hat{U}_{B}\hat{U}_{\phi }\left \vert
0\right \rangle _{a_{v}}\left \vert \psi \right \rangle ,  \label{14}
\end{equation}%
which is a form of pure state, where $\left \vert \psi \right \rangle =A\hat{%
U}_{P}\hat{U}_{F}\left \vert \alpha \right \rangle _{a}\left \vert
0\right
\rangle _{b}$.

For the case of photon loss, we can treat the system as a pure state in an
extended space. Then following Eq. (\ref{8}), we can obtain the QFI under
the pure state, denoted as $C_{Q_{a}}$, which is greater than or equal to
the QFI for mixed state, denoted as $F_{L_{a}}$, i.e., $F_{L_{a}}\leq
C_{Q_{a}}$. $C_{Q_{a}}$ is the QFI before optimizing over all possible
measurements, i.e.,%
\begin{equation}
C_{Q_{a}}=4\left[ \left \langle \psi \right \vert \hat{H}_{1}\left \vert
\psi \right \rangle -\left \vert \left \langle \psi \right \vert \hat{H}%
_{2}\left \vert \psi \right \rangle \right \vert ^{2}\right] ,  \label{15}
\end{equation}%
where $\hat{H}_{1}$ and $\hat{H}_{2}$ are defined as%
\begin{eqnarray}
\hat{H}_{1} &=&\overset{\infty }{\underset{l=0}{\sum }}\frac{d}{d\phi _{a}}%
\hat{\Pi}_{l}^{\dagger }\left( \eta ,\phi _{a},\lambda \right) \frac{d}{%
d\phi _{a}}\hat{\Pi}_{l}\left( \eta ,\phi _{a},\lambda \right) ,  \label{16}
\\
\hat{H}_{2} &=&i\overset{\infty }{\underset{l=0}{\sum }}\left[ \frac{d}{%
d\phi _{a}}\hat{\Pi}_{l}^{\dagger }\left( \eta ,\phi _{a},\lambda \right) %
\right] \hat{\Pi}_{l}\left( \eta ,\phi _{a},\lambda \right) .  \label{17}
\end{eqnarray}
Here, $\hat{\Pi}_{l}\left( \eta ,\phi _{a},\lambda \right) =\sqrt{\frac{%
\left( 1-\eta \right) ^{l}}{l!}}e^{i\phi _{a}\left( \hat{n}_{a}-\lambda
l\right) }\eta ^{\frac{\hat{n}_{a}}{2}}\hat{a}^{l}$ is the phase-dependent
Kraus operator, satisfying $\sum \hat{\Pi}_{l}^{\dagger }\left( \eta ,\phi
_{a},\lambda \right) \hat{\Pi}_{l}\left( \eta ,\phi _{a},\lambda \right) =1$%
, with $\lambda =0$ and $\lambda =-1$ representing the photon loss before
the phase shifter and after the phase shifter, respectively. $\hat{n}_{a}=%
\hat{a}^{\dag }\hat{a}$ is the number operator, and $\eta $ is related to
the dissipation factor with $\eta =1$ and $\eta =0$ corresponding to the
cases of complete lossless and absorption, respectively. Following the
spirit of Ref. \cite{h5}, we can further obtain the minimum value of $%
C_{Q_{a}}$ by optimizing over $\lambda $, corresponding to $F_{L_{a}}$,
i.e., $F_{L_{a}}=\min C_{Q_{a}}\leq C_{Q_{a}}$. Simplifying the calculation
process enables us to derive the QFI under photon loss
\begin{equation}
F_{L_{a}}=\frac{4F_{a}\eta \left \langle \hat{n}_{a}\right \rangle }{\left(
1-\eta \right) F_{a}+4\eta \left \langle \hat{n}_{a}\right \rangle },
\label{18}
\end{equation}%
where $F_{a}$ is the QFI in the ideal case \cite{h9,25}.

Next, we analyze the phase sensitivity $QCRB_{L_{a}}$\ $=1/\sqrt{F_{L_{a}}}$%
. Fig. 7 shows how $QCRB_{L_{a}}$ varies with transmittance $\eta $. From
the four sub-figures, as $\eta $ increases, $QCRB_{L_{a}}$ improves in all
cases, indicating better measurement accuracy. Overall, sensitivity improves
with increasing the order of multi-PS. For each sub-figure, different $R$
values have varying effects on sensitivity enhancement. When $R\approx 0.17$%
, the sensitivity is optimal and remains stable with $\eta $ changes,
showing minimal variation. This indicates that the FOPA system is highly
robust to photon loss, and can maintain high precision in lossy
environments, which, is thus highly valuable for practical applications at the optimal point.

\begin{figure}[tph]
\label{Fig7} \centering \includegraphics[width=0.9\columnwidth]{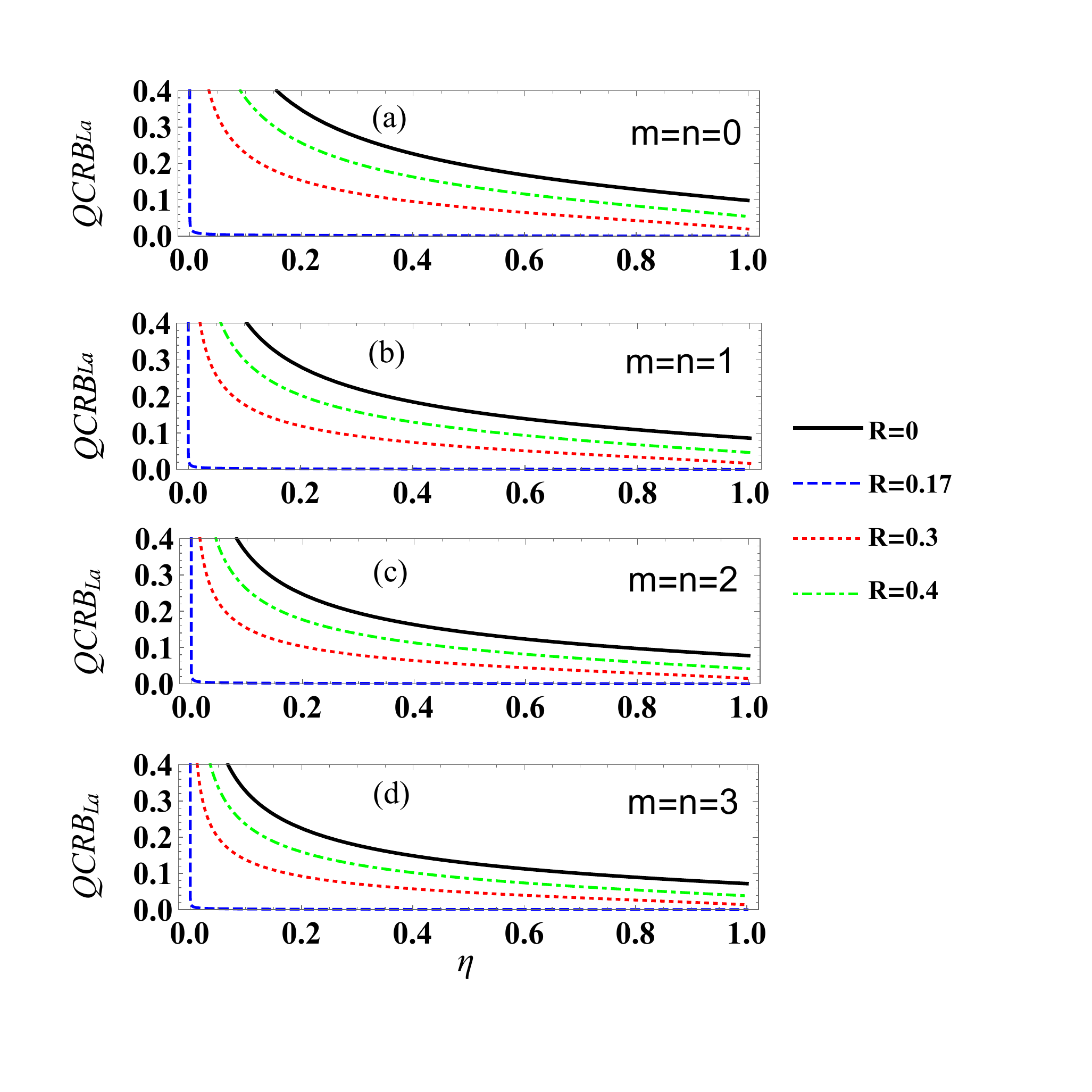}
\caption{The $QCRB_{L_{a}}$ as a function of $\protect \eta $, with $g=1$ and
$\protect \alpha =2$. }
\end{figure}

To gain a deeper understanding of how the FOPA and multi-PS defeat the
effects of different levels of photon loss, we draw Fig. 8. Looking at the
overall trend of the four sub-figures, the QCRB improves first and then
worsens as $R$ increases. And when $R\approx 0.17$, the four curves in each
sub-figure nearly overlap. This means at $R\approx 0.17$, the system is
highly robust to photon loss, which matches with the conclusion from Fig. 7.
Furthermore, comparing the sub-figures, we see that the QCRB is
significantly enhanced overall as the order of multi-PS increases.
Additionally, when $R\approx 0.17,0.3,0.4$ etc. ($R<0.5$), the multi-PS schemes ($m=n=1,2,3$)
outperform the original scheme ($m=n=0$).

\begin{figure}[tph]
\label{Fig8} \centering \includegraphics[width=0.9\columnwidth]{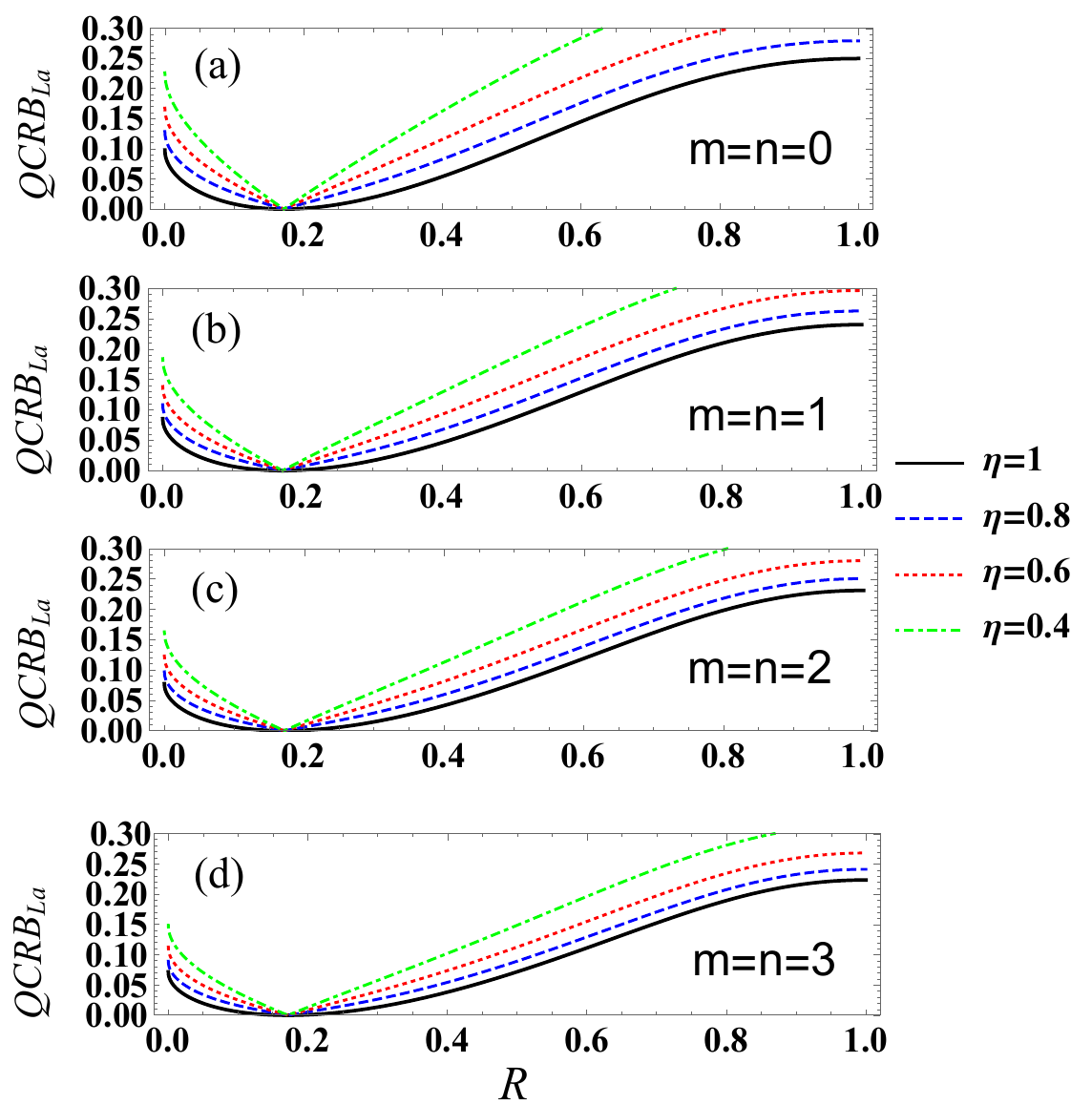}
\caption{The $QCRB_{L_{a}}$ as a function of $R$, with $g=1$ and $\protect%
\alpha =2$. }
\end{figure}

\section{\protect \bigskip Two-parameter estimation}

Current research commonly employs single-parameter QFI to analyze
single-phase estimation in interferometric systems. However, in many
practical scenarios where different phase shifts may occur in each arm, both
$\phi _{a}$\ and $\phi _{b}\ $become unknown parameters requiring
simultaneous estimation. The two-parameter problem requires the QFIM for
comprehensive phase estimation analysis \cite{h11,h10}. Additionally, in
Ref. \cite{h10}, it is indicated that in SU(1,1) interferometers,
single-parameter estimation might overestimate the precision limit. This is
because it assumes prior knowledge of certain information, like zero phase
shift in the other arm. In contrast, multiparameter estimation, which
doesn't rely on such assumptions, provides stricter and more reliable
precision bounds, effectively avoiding the overestimation issues inherent in
single-parameter estimation. Thus, in this section, we use the QFIM method
to simultaneously estimate the two phases. The two-parameter estimation
model is shown in Fig. 9.

\begin{figure}[tph]
\label{Fig9} \centering \includegraphics[width=0.9\columnwidth]{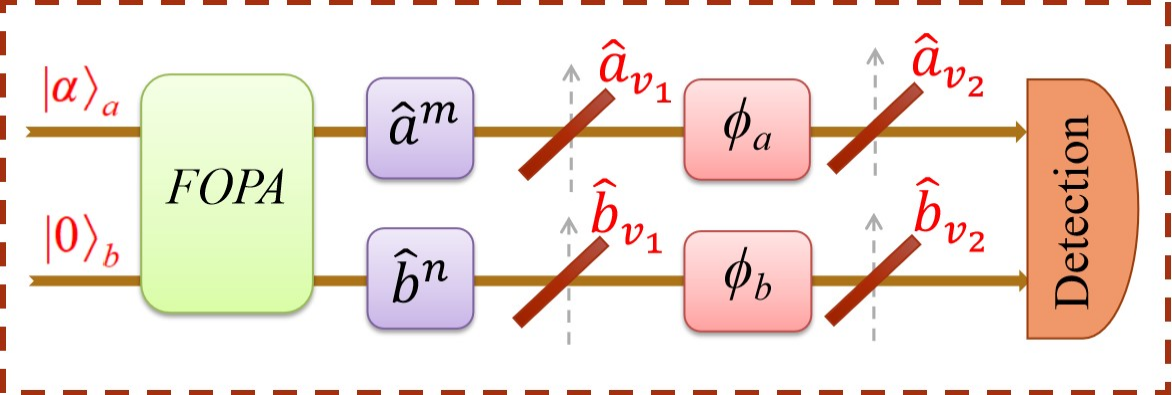}
\caption{Schematic diagram of the two-parameter estimation. The two input
ports are a coherent state $\left \vert \protect \alpha \right \rangle _{a}$
and a vacuum state $\left \vert 0\right \rangle _{b}$. $\hat{a}_{v_{1}}$, $%
\hat{a}_{v_{2}}$, $\hat{b}_{v_{1}}$, and $\hat{b}_{v_{2}}$ are vacuum modes.
$\protect \phi _{a}$ and $\protect \phi _{b}$ are the phase shifters. }
\end{figure}

\subsection{\protect \bigskip Ideal case}

\bigskip When considering two arms with unknown phase shifts, the phase
shifter operator takes the form of
\begin{equation}
\hat{U}_{\phi }=e^{-i\phi _{a}(\hat{a}^{\dagger }\hat{a})}e^{-i\phi _{b}(%
\hat{b}^{\dagger }\hat{b})},  \label{19}
\end{equation}%
with $\phi _{a}$ and $\phi _{b}$\ are the unknown phases on modes $a$\ and $%
b $, respectively. In multiparameter estimation, the QCRB is determined via
the QFIM. For the estimation of $\phi _{a}$\ and $\phi _{b}$, the QFIM is
given by a $2\times 2$\ matrix%
\begin{equation}
F_{Q}=\left[
\begin{array}{cc}
F_{aa} & F_{ab} \\
F_{ba} & F_{bb}%
\end{array}%
\right] ,  \label{20}
\end{equation}%
and
\begin{equation}
F_{ij}=4\left[ \left \langle \partial _{i}\psi _{\phi }|\partial
_{j}\psi _{\phi }\right \rangle -\left \langle \partial _{i}\psi _{\phi
}|\psi _{\phi }\right \rangle \left \langle \psi _{\phi }|\partial _{j}\psi
_{\phi }\right \rangle \right] .  \label{21}
\end{equation}%
Here $\left \vert \partial _{i}\psi _{\phi }\right \rangle =\partial
\left
\vert \psi _{\phi }\right \rangle /\partial \phi _{i}$ and $%
\left
\vert \psi _{\phi }\right \rangle =A\hat{U}_{\phi }\hat{U}_{P}\hat{U}%
_{F}\left \vert \alpha \right \rangle \left \vert 0\right \rangle $, with $%
i,j=a,b$. The diagonal elements of this matrix correspond to the QFI for the
respective parameters, while the off-diagonal entries provide information
about the parameter correlations. The QCRB for simultaneous two-phase
estimation can be derived, that is, the lower bound on the estimation
uncertainty \cite{h13,h14,h15,h16}%
\begin{equation}
\left \vert \delta \phi \right \vert ^{2}\geq \left \vert \delta \phi \right
\vert _{QCRB}^{2}=\text{Tr}\left( F^{-1}\right) .  \label{22}
\end{equation}

We analyze the $QCRB\sim R$ curve, shown in Fig. 10. Similar to
single-parameter estimation, without multi-PS operations ($m=n=0$), feedback
only improves measurement precision when $R<0.5$. However, as the order of
multi-PS increases, the measurement precision improves significantly, and
the effective range of $R$ also expands, indicating that multi-PS can
effectively enhance measurement precision. When $R$ reaches $R_{opt}$, a
singularity appears in the curve, as in previous cases, suggesting the
system may enter a special state. As shown in the local amplification of the
singularity, the curve oscillates near the singularity, indicating the
system is highly sensitive to feedback strength changes there. This
sensitivity likely stems from the feedback's impact, which can drive the
system into a critical state at specific feedback strengths. Therefore, in
practical applications, it is essential to balance measurement precision and
system stability to ensure reliability and effectiveness.

\begin{figure}[tph]
\label{Fig10} \centering \includegraphics[width=0.9\columnwidth]{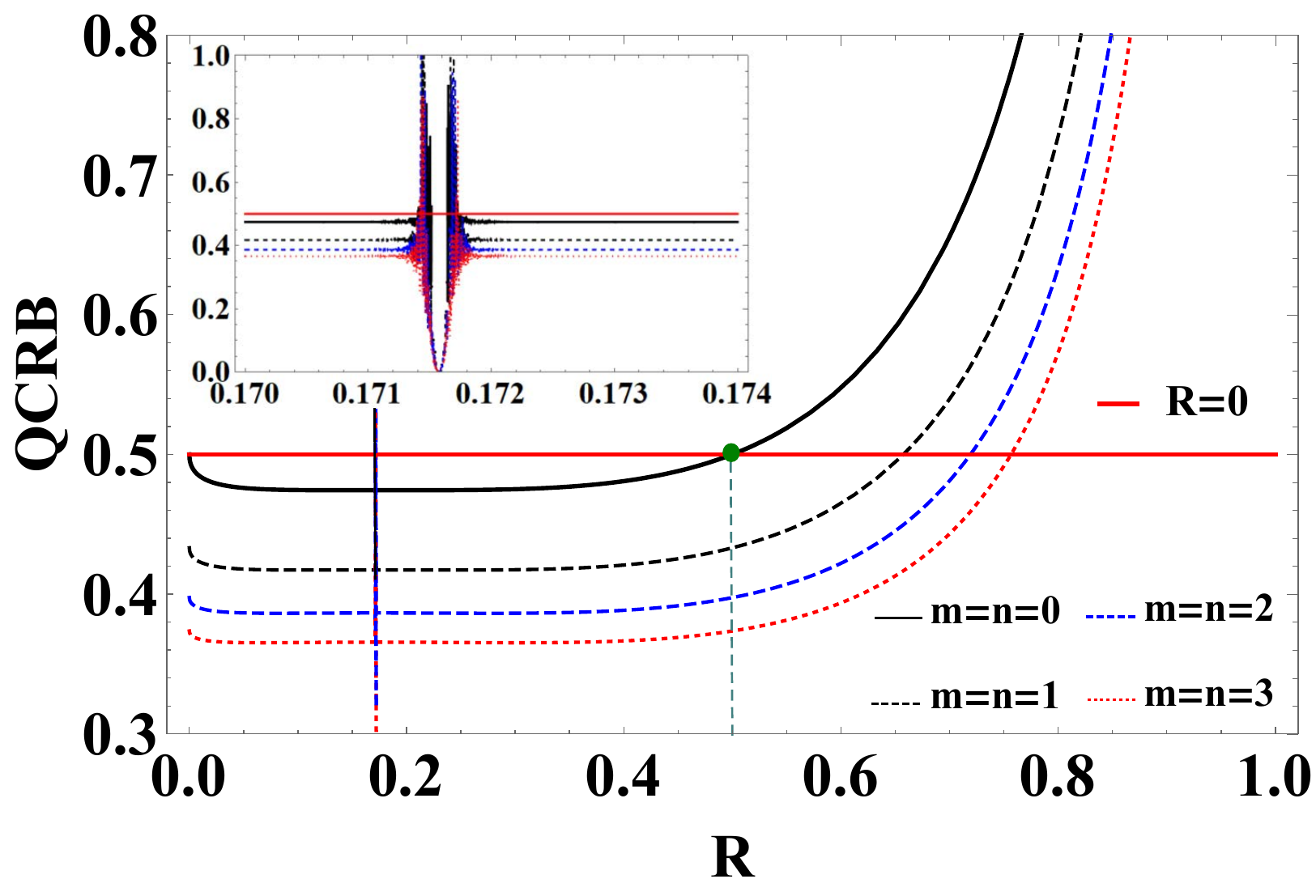}
\caption{The $QCRB$ as a function of $R$, with $g=1$ and $\protect \alpha =2$%
. The red solid line represents the sensitivity of the TOPA without multi-PS
operation, serving as a reference. $m$ and $n$ denote the order of the
multi-PS operations from mode $a$ and mode $b$, respectively.}
\end{figure}

In Figs. 11 and 12, the QCRB improves with increasing $g$ and $\alpha $,
with $R=0.17$. Our calculations show that within the improvement range of $R$
($R<0.5$), the trends for $g$ and $\alpha $ are generally consistent.
Similarly, both the FOPA and the multi-PS can enhance the measurement
precision of two-parameter estimation systems. The higher the order of
multi-PS, the better the improvement. Additionally, when compared with the
single-parameter estimation case (refer to Figs. 5 and 6): (i) For the $%
QCRB\sim g$ curve, in single-parameter estimation, as $g$ increases, the
FOPA scheme's sensitivity improvement becomes more significant. Conversely,
in two-parameter estimation, when $g$ is small, the FOPA scheme shows more
remarkable improvement. (ii) For the $QCRB\sim \alpha $ curve, in
single-parameter estimation, the FOPA scheme significantly enhances
sensitivity across the entire $\alpha $ range ($0\leq \alpha \leq 3$). By
contrast, in two-parameter estimation, the FOPA scheme only improves
sensitivity when $\alpha $ is relatively large.

\begin{figure}[tph]
\label{Fig11} \centering \includegraphics[width=0.9\columnwidth]{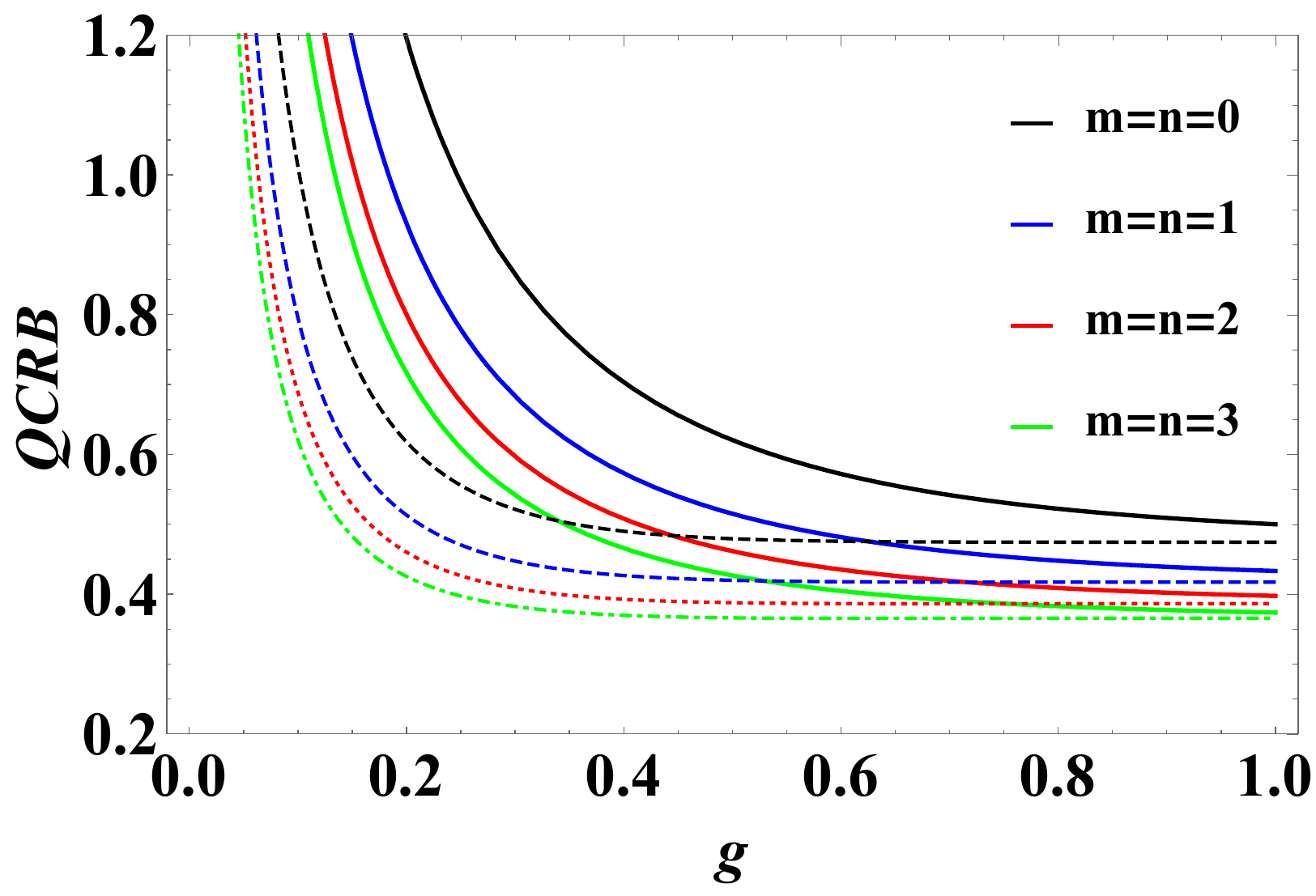}
\caption{The $QCRB$ as a function of $g$, with $R=0.17$ and $\protect \alpha %
=2$. The solid line corresponds to the system without feedback (i.e., $R=0$%
); the dashed line corresponds to the FOPA system ($R=0.17$). $m$ and $n$
denote the order of the multi-PS operations from mode $a$ and mode $b$,
respectively.}
\end{figure}

\begin{figure}[tph]
\label{Fig12} \centering \includegraphics[width=0.9\columnwidth]{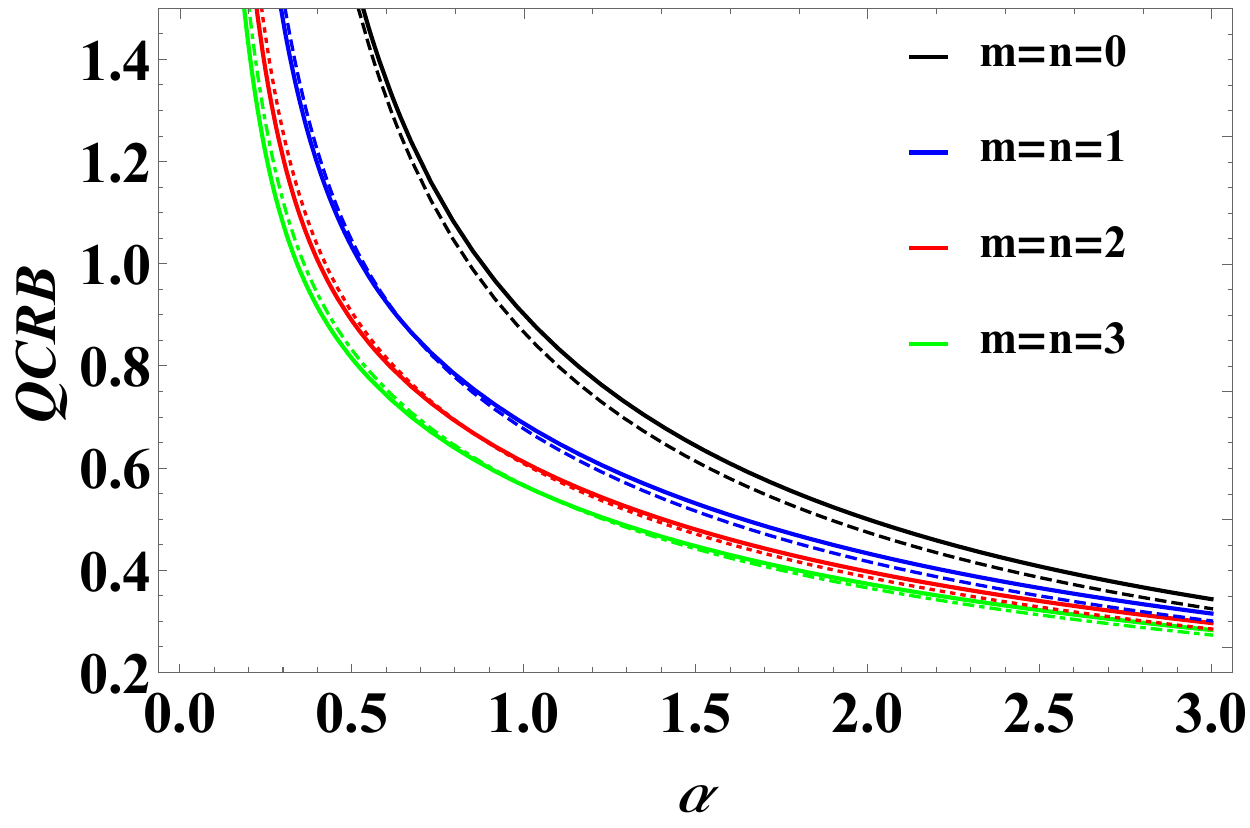}
\caption{The $QCRB$ as a function of $\protect \alpha $, with $R=0.17$ and $%
g=1$. The solid line corresponds to the system without feedback (i.e., $R=0$%
); the dashed line corresponds to the FOPA system ($R=0.17$). $m$ and $n$
denote the order of the multi-PS operations from mode $a$ and mode $b$,
respectively.}
\end{figure}

\subsection{\protect \bigskip Photon-loss case}

In this subsection, under our model, we study the simultaneous two-phase
estimation in the case of photon loss, assumed before and after the phase
shifter. Photon loss causes the probe state $\left \vert \psi \right \rangle
_{S}$ to interact with the environment $E$, leaking phase information. Due
to photon loss, the ideal two-parameter QCRB formula isn't applicable. Yue
\textit{et al.} \cite{18b} extended Escher \textit{et al}.'s variational
method \cite{h5,h18} to multi-parameter estimation and derived the QCRB for
multi-parameter estimation with photon loss. So, we briefly review this
method.

Given the initial probe state $\left \vert \psi \right \rangle _{S}$, the
phase-encoding process in a photon-loss environment is non-unitary. To
address this, we extend the Hilbert space $S$ of the probe state by
including the photon-loss environment space $E$. This ensures the probe
state undergoes unitary evolution $\hat{U}_{S+E}\left( \phi \right) $ in the
expanded system-environment space $S+E$, which can be represented by the
following equation
\begin{eqnarray}
\left \vert \psi \left( \phi \right) \right \rangle _{S+E} &=&\hat{U}%
_{S+E}\left( \phi \right) \left \vert \psi \right \rangle _{S}\left \vert
0\right \rangle _{E}  \notag \\
&=&\overset{\infty }{\underset{l}{\sum }}\hat{\Pi}_{l}\left( \phi \right)
\left \vert \psi \right \rangle _{S}\left \vert l\right \rangle _{E},
\label{23}
\end{eqnarray}%
where, $\hat{U}_{S+E}\left( \phi \right) =\hat{U}_{S_{a}+E_{a}}^{a}\left(
\phi _{a}\right) \otimes \hat{U}_{S_{b}+E_{b}}^{b}\left( \phi _{b}\right) $,
and $\left \vert 0\right \rangle _{E}=\left \vert 0\right \rangle
_{E_{a}}\otimes \left \vert 0\right \rangle _{E_{b}}$ represents the initial
state of the photon-loss environment space $E$. $\left \vert l\right \rangle
_{E}=\left \vert l_{a}\right \rangle _{E_{a}}\otimes \left \vert
l_{b}\right
\rangle _{E_{b}}$ is the orthogonal basis of the photon-loss
environment, and\ $\hat{\Pi}_{l}\left( \phi \right) =\hat{\Pi}_{l_{a}}\left(
\phi _{a}\right) \otimes \hat{\Pi}_{l_{b}}\left( \phi _{b}\right) $ is the
tensor product of Kraus operators, which are generally defined as follows%
\begin{equation}
\hat{\Pi}_{l_{k}}\left( \phi _{k}\right) =\sqrt{\frac{\left( 1-\eta
_{k}\right) ^{l_{k}}}{l_{k}!}}e^{i\phi _{k}\left( \hat{n}_{k}-\lambda
_{k}l_{k}\right) }\eta _{k}^{\frac{\hat{n}_{k}}{2}}\left( \hat{k}\right)
^{l_{k}},  \label{24}
\end{equation}%
with $k\in \left \{ a,b\right \} $, and $\eta _{k}$ quantifies the
photon-loss intensity in mode $k$, modeled via a virtual BS (see Fig. 9).
Note that $\eta _{k}=1$ ( $\eta _{k}=0$) and $\lambda _{k}=0$ ( $\lambda
_{k}=-1$) denote the ideal case (complete absorption) and photon loss
occurring before (after) the phase-encoding process, respectively.

\bigskip In this scenario, the QCRB for simultaneous two-phase estimation
under photon loss can be given by%
\begin{equation}
\left \vert \delta \phi \right \vert _{QCRB_{L}}^{2}=\max_{\hat{\Pi}%
_{l}(\phi )}\text{Tr}[C_{Q}^{-1}(\phi ,\hat{\Pi}_{l}(\phi ))].  \label{25}
\end{equation}%
Here, $C_{Q}(\phi ,\hat{\Pi}_{l}(\phi ))$ is the matrix of the QFIM in the
extended system space $S+E$, in which the matrix elements can be given by
\begin{equation}
C_{Q_{jk}}(\phi ,\hat{\Pi}_{l}(\phi ))=4\left[ \left \langle \hat{\Upsilon}%
_{jk}\right \rangle -\left \langle \hat{\Xi}_{j}\right \rangle \left \langle
\hat{\Xi}_{k}\right \rangle \right] ,  \label{26}
\end{equation}%
where $j,k\in \left \{ a,b\right \} $, and $\left \langle \cdot
\right
\rangle $ denotes the average over the probe state $\left \vert \psi
\right
\rangle _{S}$, with
\begin{eqnarray}
\hat{\Xi}_{j} &=&i\overset{\infty }{\underset{l_{j}}{\sum }}\frac{d\hat{\Pi}%
_{l_{j}}^{\dagger }\left( \phi _{j}\right) }{d\phi _{j}}\hat{\Pi}%
_{l_{j}}\left( \phi _{j}\right) ,  \notag \\
\hat{\Upsilon}_{jk} &=&\left \{
\begin{array}{c}
\overset{\infty }{\underset{l_{j}}{\sum }}\frac{d\hat{\Pi}_{l_{j}}^{\dagger
}\left( \phi _{j}\right) }{d\phi _{j}}\frac{d\hat{\Pi}_{l_{j}}\left( \phi
_{j}\right) }{d\phi _{j}},\text{ }j=k, \\
\hat{\Xi}_{j}\hat{\Xi}_{k},\text{ }j\neq k.%
\end{array}%
\right.  \label{27}
\end{eqnarray}%
Based on the Kraus operator, Eq. (\ref{27}) can be simplified to be \cite%
{h20}%
\begin{align}
\hat{\Xi}_{j}& =\mu _{j}\hat{n}_{j},  \notag \\
\hat{\Upsilon}_{jk}& =\left \{
\begin{array}{c}
\mu _{j}^{2}\hat{n}_{j}^{2}+\gamma _{j}\hat{n}_{j},\text{ }j=k, \\
\\
\hat{\Xi}_{j}\hat{\Xi}_{k},\text{ }j\neq k,%
\end{array}%
\right.  \label{28}
\end{align}%
where $\mu _{j}=1-(1-\eta _{j})(1+\lambda _{j}),\gamma _{j}=\eta _{j}(1-\eta
_{j})(1+\lambda _{j})^{2}$.

For ease of discussion, we assume $\eta _{j}=\eta $ and $\lambda
_{j}=\lambda $. According to Eqs. (\ref{25}) and (\ref{28}), we can derive
the QCRB lower bounds in a photon-loss environment. Note that $\lambda $ is
a variational parameter. Finally, a numerical calculation is used to
maximize Tr$[C_{Q}^{-1}(\phi ,\hat{\Pi}_{l}(\phi ))]$, which gives the QCRB
under photon loss.

\begin{figure}[tph]
\label{Fig13} \centering \includegraphics[width=0.9\columnwidth]{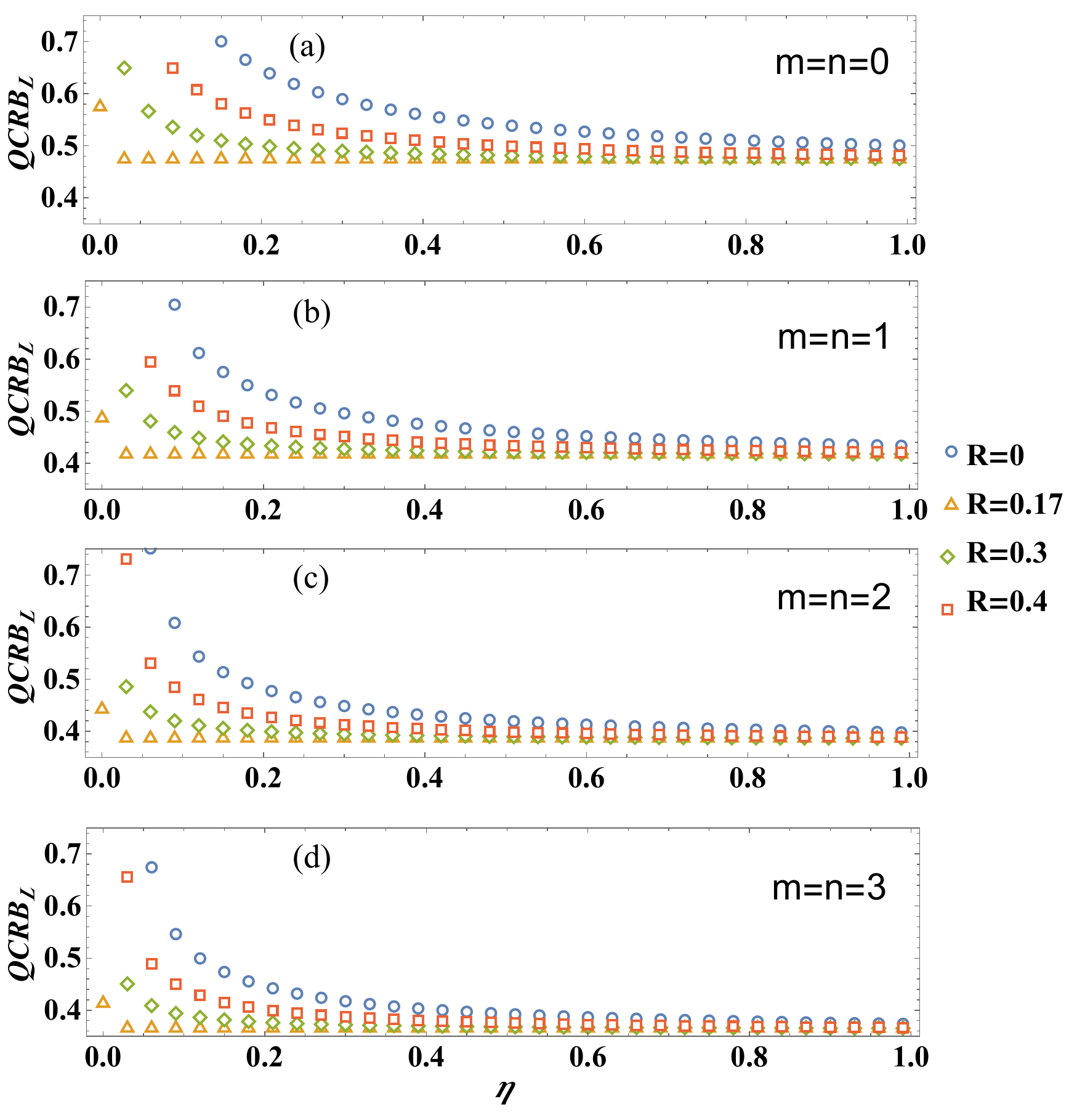}
\caption{The $QCRB_{L}$ as a function of $\protect \eta $, with $g=1$ and $%
\protect \alpha =2$. }
\end{figure}

As illustrated in Fig. 13, $QCRB_{L}$ shows enhancement with an increase in $%
\eta $, signaling a rise in measurement precision. Furthermore, by comparing
the four sub-figures, we can find that the sensitivity of the original
system ($m=n=0$) can be further enhanced after undergoing the multi-PS
operations ($m=n=1,2,3$), and the higher the order of the multi-PS, the
better the improvement. In each sub-figure, sensitivity enhancement varies
with different $R$ values. When $R\approx 0.17$, sensitivity peaks and
remains stable despite $\eta $ fluctuations, demonstrating minimal
deviation. It implies the FOPA system is highly robust to photon loss,
capable of preserving measurement accuracy in lossy settings, and thus holds
significant promise for real-world applications.

In Fig. 14, analyzing the collective trend across the four sub-figures
reveals that the $QCRB_{L}$ initially improves with increasing R, reaches an
optimal point, and then deteriorates. Notably, at $R\approx 0.17$, the
curves in each sub-figure almost converge. This convergence implies that the
system is particularly robust against photon loss at this $R$ value, a
finding that aligns with the results from Fig. 13. Overall, the results
further confirm that the introduction of the multi-PS operations (with $%
m=n=1,2,3$) can enhance the sensitivity of the original system ($m=n=0$).
The improvement of sensitivity becomes increasingly significant with the
higher order of the multi-PS operation. These conclusions are consistent
with the single-parameter estimation.

\begin{figure}[tph]
\label{Fig14} \centering \includegraphics[width=0.9\columnwidth]{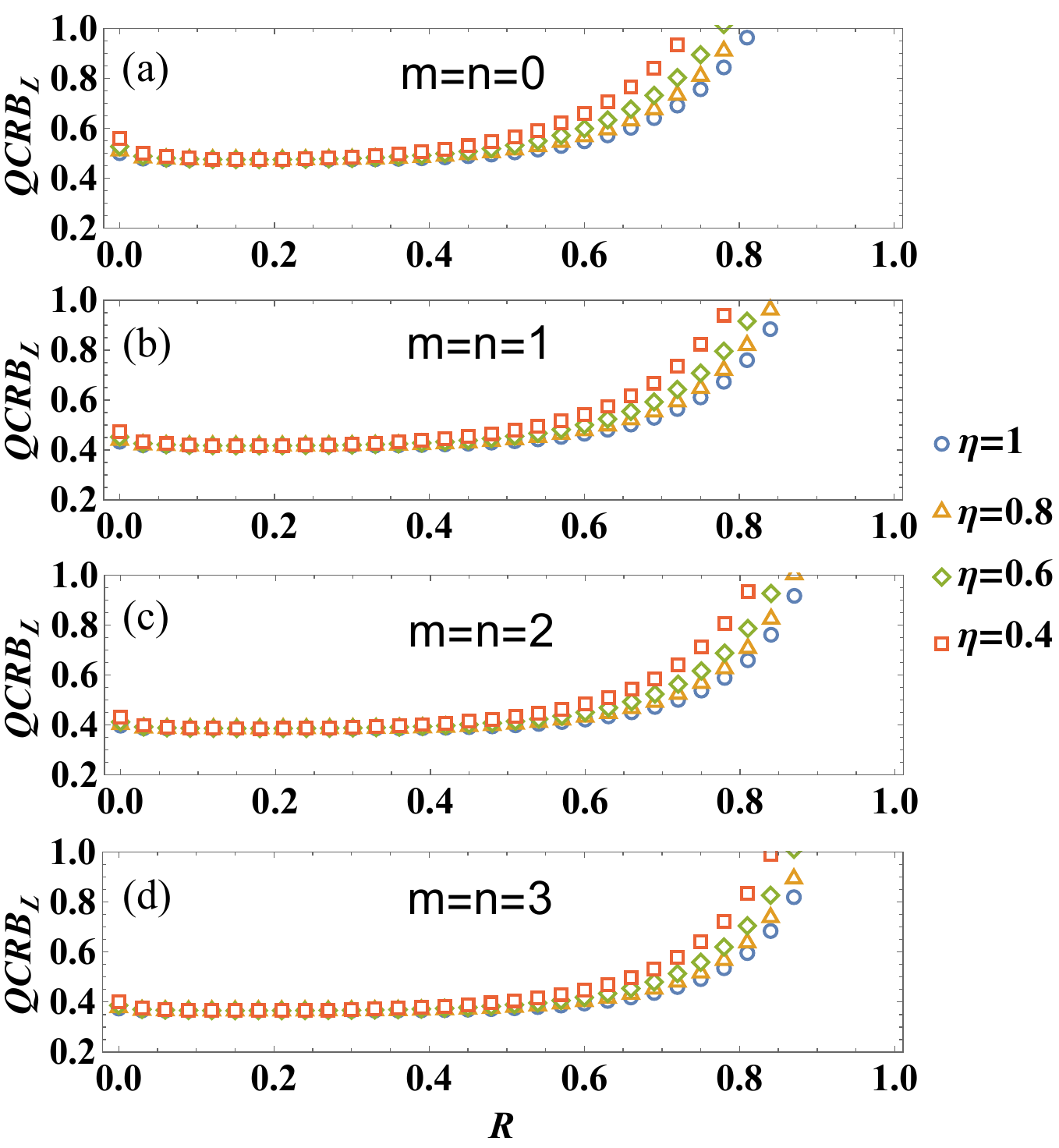}
\caption{The $QCRB_{L}$ as a function of $R$, with $g=1$ and $\protect \alpha %
=2$. }
\end{figure}

\section{Discussions: Intramode correlation and intermode correlation}

In order to delve deeper into how the FOPA and multi-PS operations boost
phase measurement precision, we leverage second-order coherence functions to
quantify the simultaneous two-parameter estimation. Specifically, we examine
the intramode correlation functions $g_{a}^{(2)}=\langle \hat{a}^{\dagger 2}%
\hat{a}^{2}\rangle /\langle \hat{a}^{\dagger }\hat{a}\rangle ^{2}$, $%
g_{b}^{(2)}=\langle \hat{b}^{\dagger 2}\hat{b}^{2}\rangle /\langle \hat{b}%
^{\dagger }\hat{b}\rangle ^{2}$, and the intermode correlation function $%
g_{ab}^{(2)}=\langle \hat{a}^{\dagger }\hat{a}\hat{b}^{\dagger }\hat{b}%
\rangle /(\langle \hat{a}^{\dagger }\hat{a}\rangle \langle \hat{b}^{\dagger }%
\hat{b}\rangle )$ \cite{h21}. These functions help us analyze the QCRB's
dependence on both intramode and intermode correlations. Within photonic
coherence theory, $g_{a}^{(2)}$\ and $g_{b}^{(2)}$ correspond to the
probability of detecting two photons simultaneously at detectors $A$ and $B$%
, respectively. Meanwhile, $g_{a}^{(2)}$\ and $g_{b}^{(2)}$ reflect the
likelihood of detecting a photon at detector $A$ and another at detector $B$
simultaneously, as depicted in Fig. 15.

\begin{figure}[tph]
\label{Fig15} \centering \includegraphics[width=0.9\columnwidth]{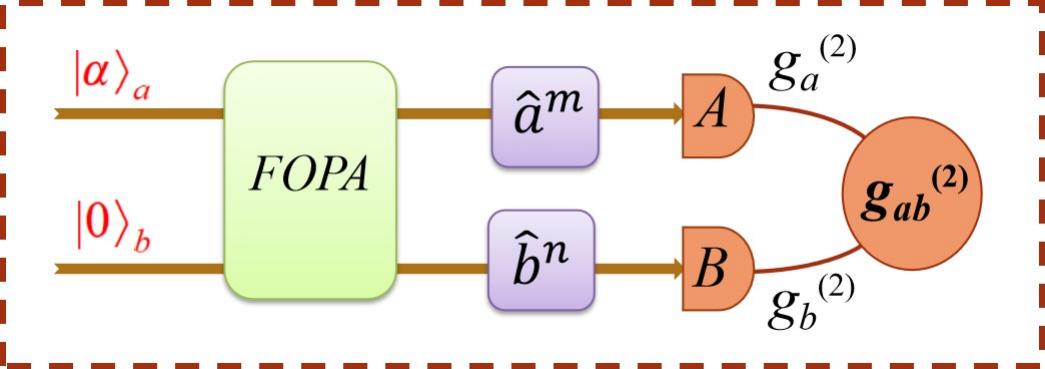}
\caption{The sketch of the distinction between the intramode correlations $%
g_{a}^{(2)}$ and $g_{b}^{(2)}$ and the intermode correlation $g_{ab}^{(2)}$.
}
\end{figure}

\begin{figure}[tph]
\label{Fig16} \centering \includegraphics[width=0.9\columnwidth]{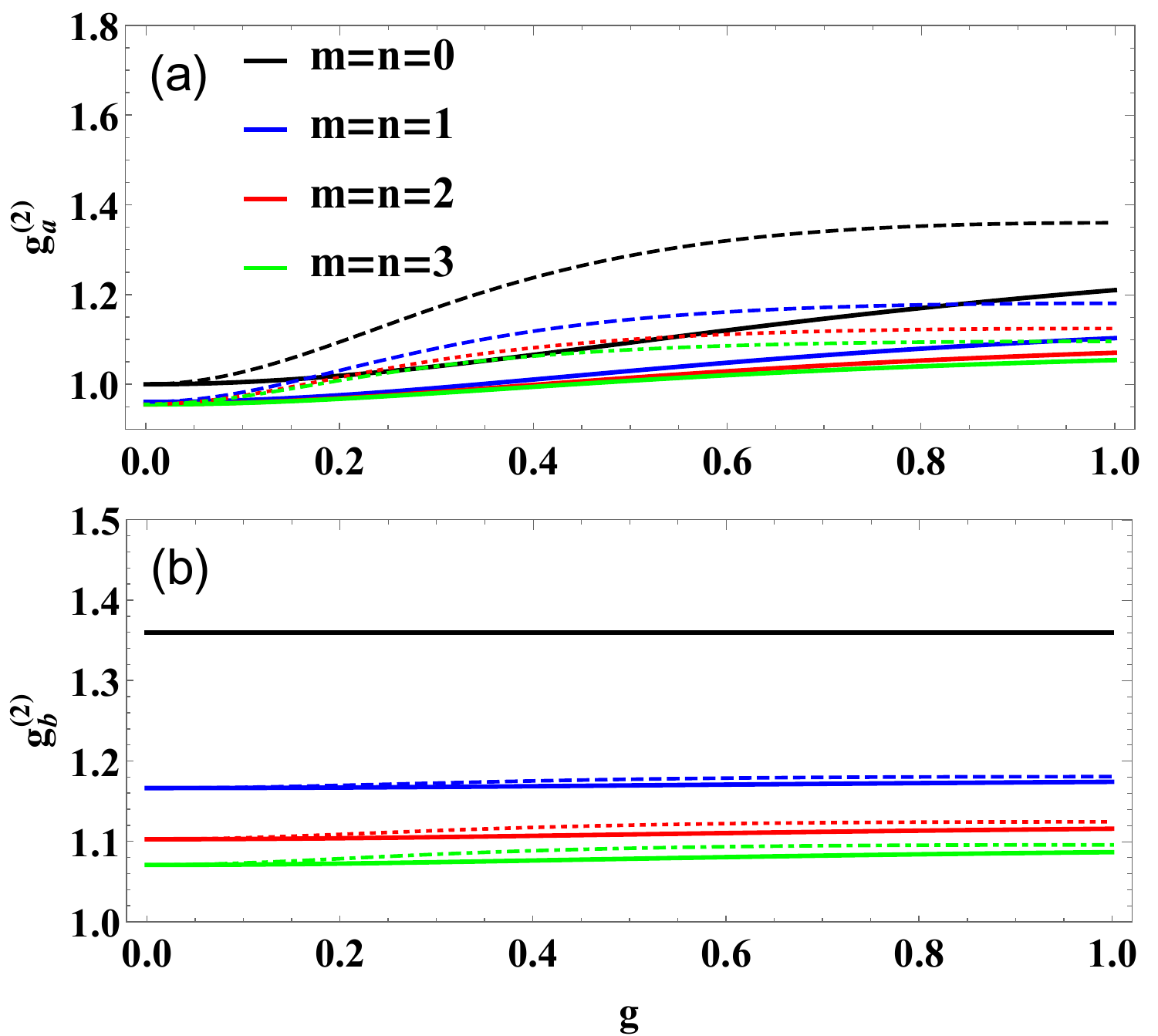}
\caption{Both the intarmode correlation function (a) $g_{a}^{(2)}$ and (b) $%
g_{b}^{(2)}$ as a function of $g$, with $\protect \alpha =2$. The solid line
corresponds to the system without feedback (i.e., $R=0$); the dashed line
corresponds to the FOPA system ($R=0.17$). $m$ and $n$ denote the order of
the multi-PS operations from mode $a$ and mode $b$, respectively.}
\end{figure}

\begin{figure}[tph]
\label{Fig17} \centering \includegraphics[width=0.9\columnwidth]{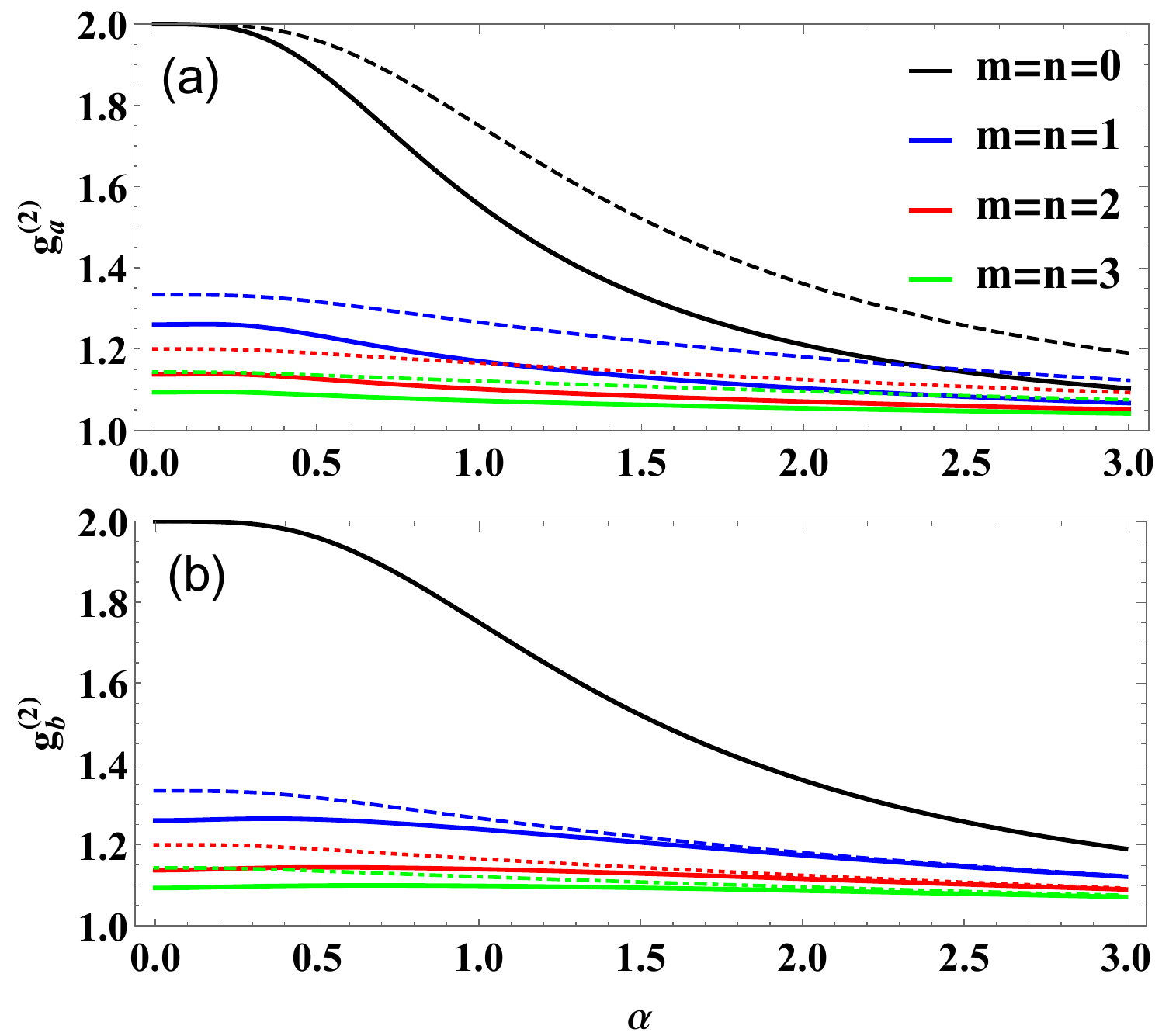}
\caption{Both the intarmode correlation function (a) $g_{a}^{(2)}$ and $%
g_{b}^{(2)}$ as a function of $\protect \alpha $, with $g=1$. The solid line
corresponds to the system without feedback (i.e., $R=0$); the dashed line
corresponds to the FOPA system ($R=0.17$). $m$ and $n$ denote the order of
the multi-PS operations from mode $a$ and mode $b$, respectively.}
\end{figure}

From Figs. 16 and 17, we can observe that modes $a$ and $b$ exhibit
different intramode correlation characteristics. $g_{a}^{(2)}$ shows an
upward trend as $g$ increases (see Fig. 16(a)), while $g_{b}^{(2)}$ remains
largely unchanged with varying $g$ (see Fig. 16(b)). As $\alpha $ increases,
both $g_{a}^{(2)}$ and $g_{b}^{(2)}$ decrease (refer to Fig. 17).
Specifically, when $m=n=0$, the FOPA does not enhance $g_{b}^{(2)}$,
aligning with the no-feedback system's effect. In Fig. 18, the intermode
correlation $g_{ab}^{(2)}$ decreases with increasing $g$ and $\alpha $, but
is more sensitive to $\alpha $ changes. By comparing the intramode
correlation and intermode correlation curves, we can observe that: (i) FOPA
shows higher intramode and lower intermode correlations than TOPA, thereby
enhancing phase-measurement precision. (ii) As the order of multi-PS
increases, although intramode correlation is somewhat weakened, intermode
correlation drops significantly. Even though multi-PS doesn't boost
intramode correlation like Gaussian operations, its greater reduction in
intermode correlation positively affects phase sensitivity. Overall, these
effects improve phase sensitivity.

\begin{figure}[tph]
\label{Fig18} \centering \includegraphics[width=0.87\columnwidth]{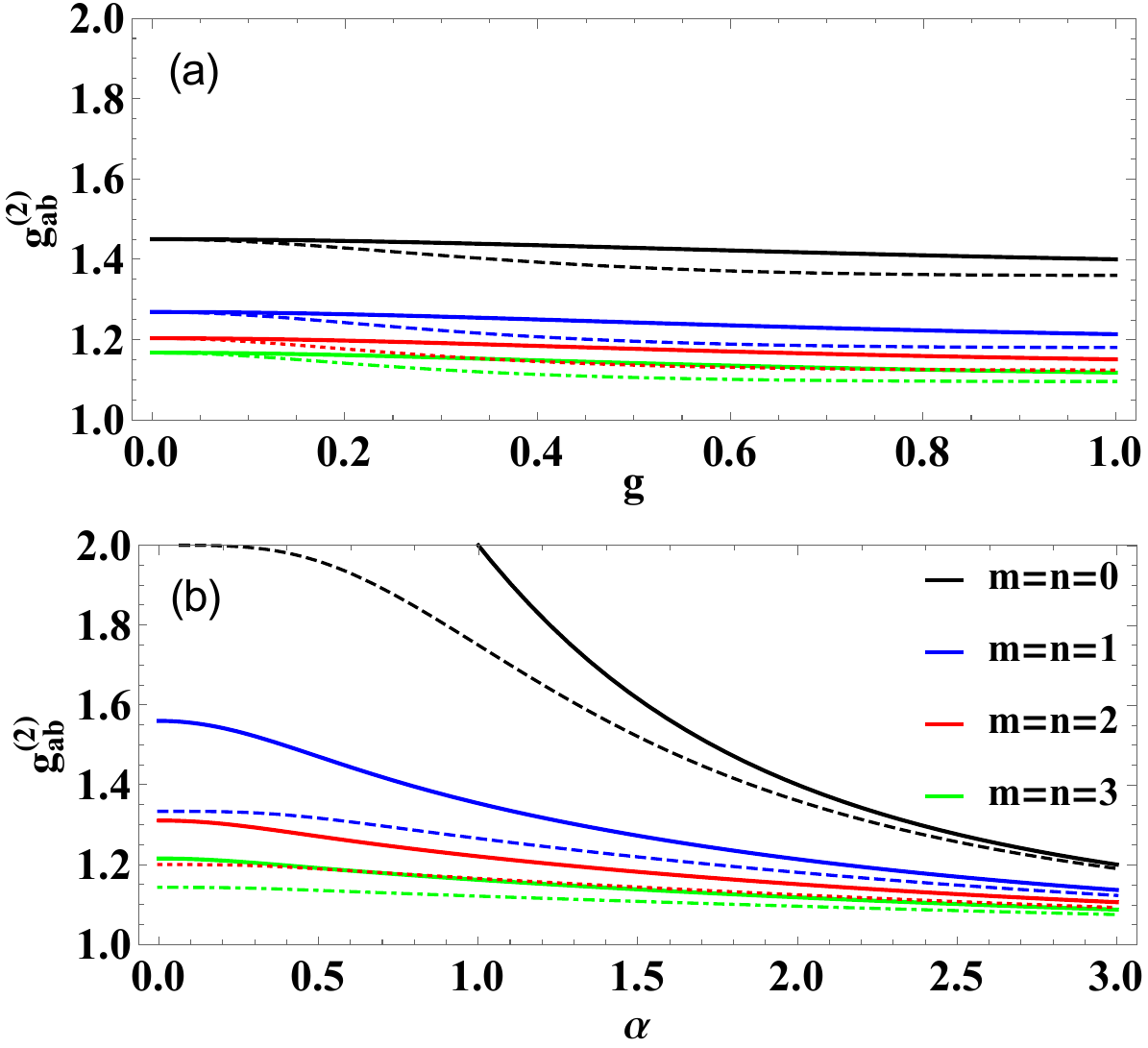}
\caption{(a) The intermode correlation function $g_{ab}^{(2)}$ as a function
of $g$, with $\protect \alpha =2$. (b) The intermode correlation function $%
g_{ab}^{(2)}$ as a function of $\protect \alpha $, with $g=1$. The solid line
corresponds to the system without feedback (i.e., $R=0$); the dashed line
corresponds to the FOPA system ($R=0.17$). $m$ and $n$ denote the order of
the multi-PS operations from mode $a$ and mode $b$, respectively.}
\end{figure}

To gain a deeper understanding of how the FOPA affects intramode and
intermode correlations, we plot the variation of correlations with $R$ under different schemes in Fig. 19.
It reveals that when $R<0.5$, the
FOPA scheme enhances intramode correlation while reducing intermode
correlation. This aligns closely with the prior conclusion from Fig. 3 (the
FOPA can improve the phase sensitivity when $R<0.5$). Thus, for simultaneous
two-phase estimation, increasing intramode correlation while decreasing
intermode correlation may improve estimation accuracy, a finding consistent
with references \cite{h20,h21}.

\begin{figure}[tph]
\label{Fig19} \centering \includegraphics[width=0.9\columnwidth]{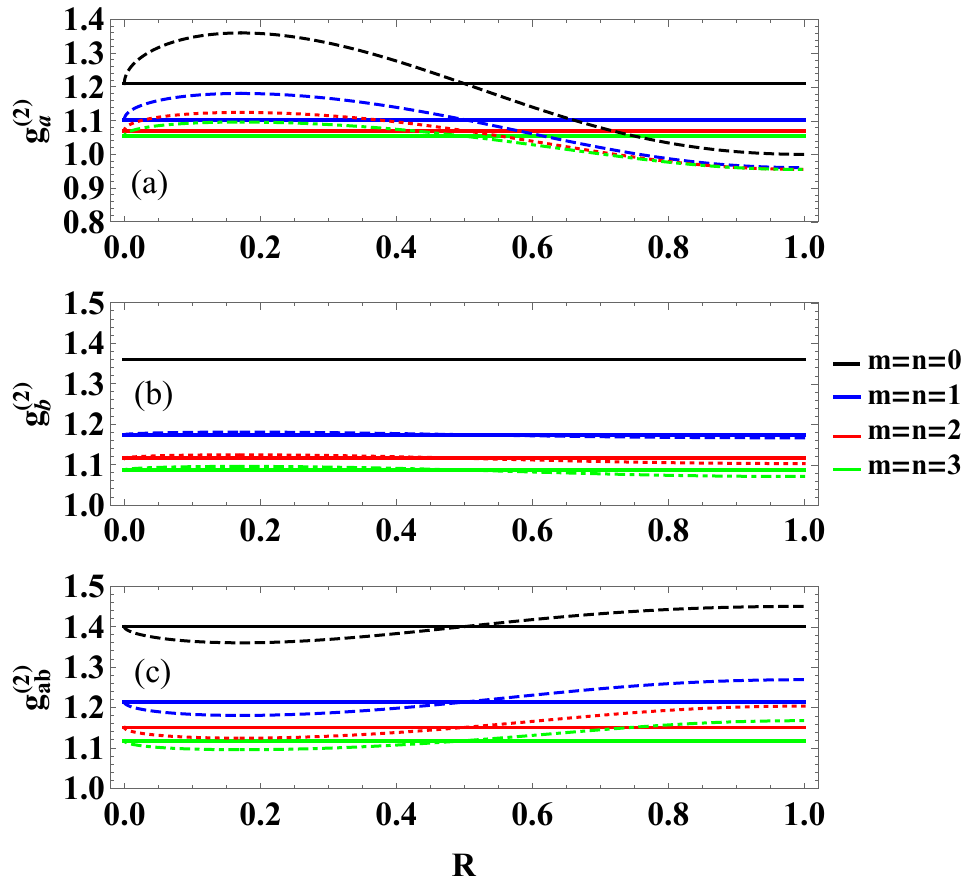}
\caption{(a) The intramode correlation function $g_{a}^{(2)}$, (b) the
intramode correlation function $g_{b}^{(2)}$ and (c) the intermode
correlation function $g_{ab}^{(2)}$ as functions of $R$, with $g=1$ and $%
\protect \alpha =2$. The solid line corresponds to the system without
feedback (i.e., $R=0$); the dashed line corresponds to the FOPA system. $m$
and $n$ denote the order of the multi-PS operations from mode $a$ and mode $b
$, respectively.}
\end{figure}

\section{Conclusion}

In this paper, we analyze how multi-PS in the FOPA-assisted interferometer
enhances measurement precision for single-parameter and two-parameter
estimation, considering both ideal and photon-loss cases. We also examine
the effects of the FOPA's feedback strength $R$, OPA's gain $g$, coherent
state amplitude $\alpha $, and multi-PS's order on system performance. An
optimal feedback strength $R_{opt}$ exists for both estimation scenarios.
Selecting a suitable $R$ can significantly enhance robustness to photon loss
and improve measurement precision. The introduced FOPA and multi-PS
operations effectively enhance estimation precision in both single-parameter
and two-parameter cases. Additionally, our analysis of QCRB's dependence on
intramode correlations ($g_{a}^{(2)}$ and $g_{b}^{(2)}$) and intermode
correlation ($g_{ab}^{(2)}$) reveals that boosting intramode while reducing
intermode correlations may be beneficial for estimation accuracy.

Theoretically, this study combines the feedback control and non-Gaussian
operations, offering novel methods for multiparameter estimation.
Practically, it can optimize parameter estimation precision in
quantum-enhanced measurements, enhance system stability, and provide a
foundation for high-precision quantum measurements.

Future research could explore expanding the analysis to more complex
multiparameter estimation scenarios, further exploring the potential of
non-Gaussian operations, developing more efficient feedback control
strategies, and conducting experimental research to verify and refine the
theoretical findings.

\begin{acknowledgments}
This work is supported by the National Natural Science Foundation of China (Grants No. 11964013 and No. 12104195), the Jiangxi Provincial Natural Science Foundation (Grants No. 20242BAB26009 and 20232BAB211033), as well as the Jiangxi Provincial Key Laboratory of Advanced Electronic Materials and Devices (Grant No. 2024SSY03011), Jiangxi Civil-Military Integration Research Institute (Grant No. 2024JXRH0Y07), and the Science and Technology Project of Jiangxi Provincial Department of Science and Technology (Grant No. GJJ2404102).
\end{acknowledgments}\bigskip

\textbf{APPENDIX A: THE INPUT-OUTPUT RELATION OF THE FOPA}

\bigskip

The signal flow graph method, commonly used in control theory, is adopted
here to calculate the input-output relationship of the FOPA. First, the
signal flow graph of FOPA is constructed based on its model, as shown in
Fig. 20. This signal flow graph encompasses every component of the FOPA,
representing the entire system as a network of multiple nodes and branches.
Here, nodes denote the probe states of the system, while branches represent
the system's transmission characteristics. The signal flow graph intuitively
illustrates the flow and interrelationships of probe states among different
system components, laying the foundation for the subsequent application of
Mason's formula to calculate the transfer function \cite{h22}.

\begin{figure}[tph]
\label{Fig20} \centering \includegraphics[width=0.85\columnwidth]{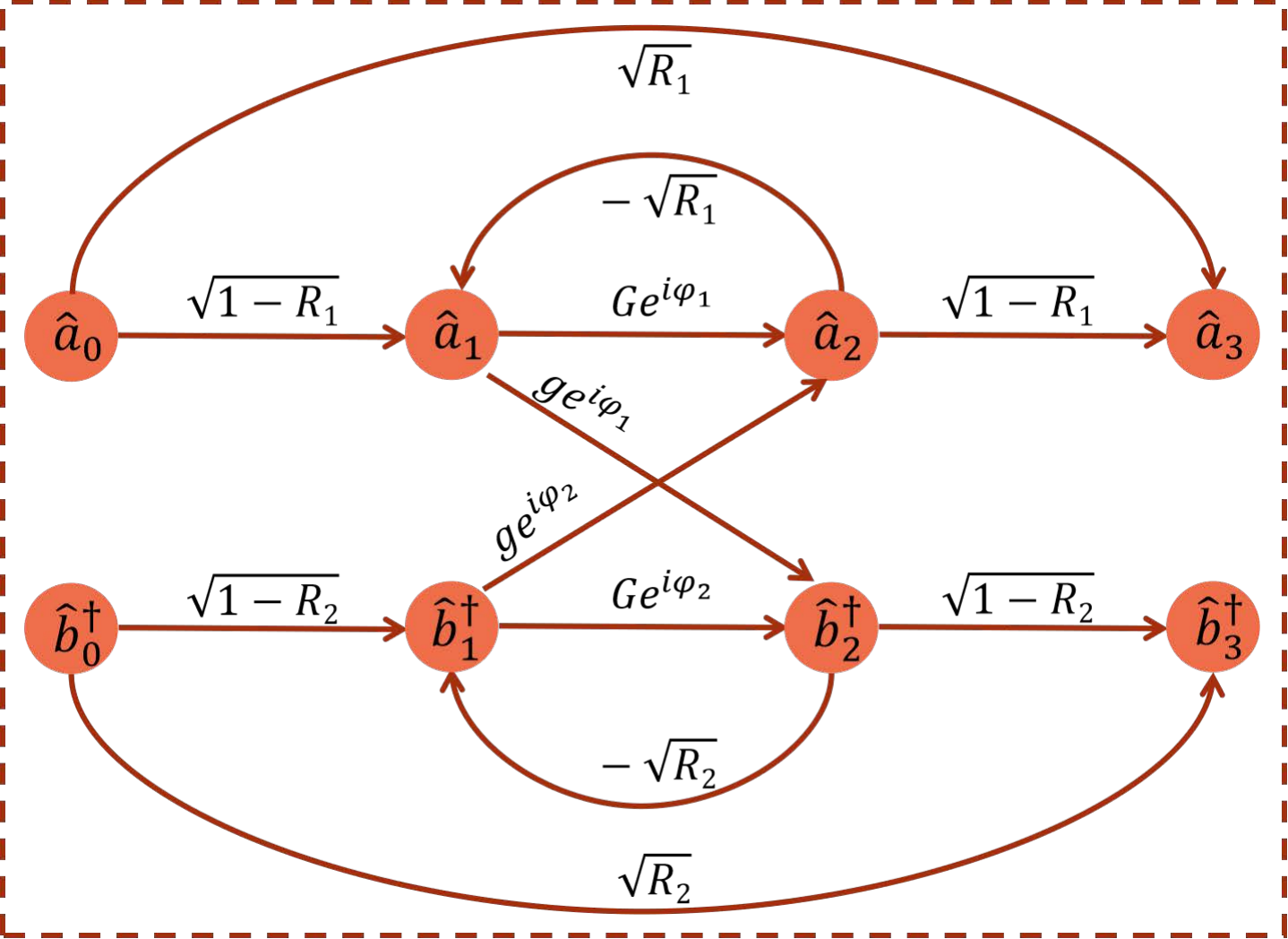}
\caption{The signal flow graph of the FOPA.}
\end{figure}

Mason's gain formula, used to determine the transfer relationship between
system input and output, takes the general form \cite{h23}%
\begin{equation}
H=\frac{\sum_{i=1}^{n}P_{i}\Delta _{i}}{\Delta },  \tag{A1}
\end{equation}%
where $i=1,2,\ldots ,n$. $\Delta $ is the determinant of the graph,
indicating the impact of all feedback loops in the system. $P_{i}$ is the
gain of the $i$-th forward path, which is the total gain of the signal
transmission along that path from the input to the output node. $\Delta _{i}$%
, the cofator of determinant for the $i$-th forward path, relates to the
path factors of all closed loops not touching the $i$-th forward path. $%
\Delta $ is given by
\begin{equation}
\Delta =1-\sum L_{a}+\sum L_{b}-\sum L_{c}+\sum L_{d}-\cdots .  \tag{A2}
\end{equation}%
In the formula, $L_{a}$ is the sum of all single loops, $L_{b}$ is the sum
of the products of all pairs of non-touching loops, $L_{c}$ is the sum of
the products of all triples of non-touching loops, and $L_{d}$ is the sum of
the products of all quadruples of non-touching loops in the signal flow
graph, and so on. Based on the signal flow graph of the FOPA, setting $%
R_{1}=R_{2}=R$ and $\varphi _{1}=\varphi _{2}=\pi $, we can obtain all the
loops: $L_{1}=G\sqrt{R},L_{2}=G\sqrt{R},L_{3}=Rg^{2}$. Accordingly, the
determinant of the transfer function for the FOPA is given by
\begin{equation}
\Delta =1-\left( L_{1}+L_{2}+L_{3}\right) +\left( L_{1}L_{2}\right) .
\tag{A3}
\end{equation}

In the transfer relationship between $\hat{a}_{3}$ and $\hat{a}_{0}$, the
forward paths and cofactors are as follows
\begin{align}
P_{1}& =\sqrt{R},\Delta _{1}=1-\left( L_{1}+L_{2}+L_{3}\right) +\left(
L_{1}L_{2}\right) .  \notag \\
P_{2}& =-(1-R)G,\Delta _{2}=1-L_{2}.  \notag \\
P_{3}& =-(1-R)\sqrt{R}g^{2},\Delta _{3}=1.  \tag{A4}
\end{align}

In the transfer relationship between $\hat{a}_{3}$ and $\hat{b}_{0}^{\dagger
}$, the forward path and cofactor are as follows
\begin{equation}
P_{1}=-g\left( 1-R\right) ,\Delta _{1}=1.  \tag{A5}
\end{equation}%
And, in the transfer relationship between $\hat{b}_{3}^{\dagger }$ and $\hat{%
b}_{0}^{\dagger }$, the forward path and cofactor are as follows%
\begin{align}
P_{1}& =\sqrt{R},\Delta _{1}=1-\left( L_{1}+L_{2}+L_{3}\right) +\left(
L_{1}L_{2}\right) .  \notag \\
P_{2}& =-(1-R)G,\Delta _{2}=1-L_{2}.  \notag \\
P_{3}& =-(1-R)\sqrt{R}g^{2},\Delta _{3}=1.  \tag{A6}
\end{align}%
In the transfer relationship between In the transfer relationship between $%
\hat{b}_{3}^{\dagger }$ and $\hat{a}_{0}$, the forward path and cofactor are
as follows%
\begin{equation}
P_{1}=-g\left( 1-R\right) ,\Delta _{1}=1.  \tag{A7}
\end{equation}%
Substituting Eqs. (A3)$\thicksim $(A7) into Eq. (A1) gives the input-output
relationship of the FOPA, which is shown in Eq. (\ref{4}).

\bigskip \textbf{APPENDIX B: THE CALCULATION OF TWO-PARAMETER ESTIMATION FOR
THE FOPA}

\bigskip Before calculating the QCRB, we first employ the transformation
relation of the FOPA in Eq. (\ref{4}) to calculate the following general
formula \cite{h24},%
\begin{align}
& \Gamma _{^{m,n,x_{1},y_{1},x_{2},y_{2}}}  \notag \\
& =\left \langle 0\right \vert \left \langle \alpha \right \vert \hat{U}%
_{F}^{\dagger }\hat{U}_{P}^{\dagger }\left( \hat{a}^{\dagger x_{1}}\hat{a}%
^{y_{1}}\hat{b}^{\dagger x_{2}}\hat{b}^{y_{2}}\right) \hat{U}_{P}\hat{U}%
_{F}\left \vert \alpha \right \rangle \left \vert 0\right \rangle  \notag \\
& =\frac{\partial ^{x_{1}+y_{1}+x_{2}+y_{2}+2m+2n}}{\partial \lambda
_{1}^{m+x_{1}}\partial \lambda _{2}^{m+y_{1}}\partial \lambda
_{3}^{x_{2}+n}\partial \lambda _{4}^{y_{2}+n}}e^{\left( \lambda _{1}\frac{%
k_{1}^{\ast }}{k_{0}^{\ast }}+\lambda _{4}\frac{k_{4}^{\ast }}{k_{0}^{\ast }}%
\right) \alpha ^{\newline
\ast }}  \notag \\
& \times e^{\lambda _{4}\frac{k_{4}^{\ast }}{k_{0}^{\ast }}\left( \lambda
_{2}\frac{k_{1}}{k_{0}}+\lambda _{3}\frac{k_{4}}{k_{0}}\right) +\lambda _{1}%
\frac{k_{3}^{\ast }}{k_{0}^{\ast }}\left( \lambda _{2}\frac{k_{3}}{k_{0}}%
+\lambda _{3}\frac{k_{2}}{k_{0}}\right) }  \notag \\
& \times e^{\left( \lambda _{2}\frac{k_{1}}{k_{0}}+\lambda _{3}\frac{k_{4}}{%
k_{0}}\right) \alpha }|_{\lambda _{1}=\lambda _{2}=\lambda _{3}=\lambda
_{4}=0}.  \tag{B1}
\end{align}%
Here, $m$, $n$, $x_{1}$, $y_{1}$, $x_{2}$ and $y_{2}$ are positive integers.
$\lambda _{1}$, $\lambda _{2}$, $\lambda _{3}$ and $\lambda _{4}$ are
differential variables. After differentiation, all these differential
variables take zero.

Thus, the QCRB in ideal case is%
\begin{equation}
QCRB=\sqrt{\frac{F_{aa}+F_{bb}}{F_{aa}F_{bb}-F_{ab}F_{ba}}},  \tag{B2}
\end{equation}%
where%
\begin{align}
F_{aa}& =4[A^{2}\Gamma _{m,n,2,2,0,0}+A^{2}\Gamma _{m,n,1,1,0,0}  \notag \\
& -\left( A^{2}\Gamma _{m,n,1,1,0,0}\right) ^{2}],  \tag{B3}
\end{align}%
and%
\begin{align}
F_{ab}& =4[A^{2}\Gamma _{m,n,1,1,1,1}  \notag \\
& -\left( A^{2}\Gamma _{m,n,1,1,0,0}\right) \left( A^{2}\Gamma
_{m,n,0,0,1,1}\right) ],  \tag{B4}
\end{align}%
\begin{equation}
F_{ba}=F_{ab},  \tag{B5}
\end{equation}%
as well as
\begin{align}
F_{bb}& =4[A^{2}\Gamma _{m,n,0,0,2,2}+A^{2}\Gamma _{m,m,0,0,1,1}  \notag \\
& -\left( A^{2}\Gamma _{m,n,0,0,1,1}\right) ^{2}].  \tag{B6}
\end{align}

The QCRB under loss conditions is%
\begin{equation}
QCRB_{L}=\sqrt{\max \frac{C_{Q_{22}}+C_{Q_{11}}}{%
C_{Q_{11}}C_{Q_{22}}-C_{Q_{12}}C_{Q_{21}}}},  \tag{B7}
\end{equation}
where
\begin{align}
C_{Q_{11}}& =4\{ \mu ^{2}\left( A^{2}\Gamma _{m,n,2,2,0,0}+A^{2}\Gamma
_{m,n,1,1,0,0}\right)  \notag \\
& +\gamma \left( A^{2}\Gamma _{m,n,1,1,0,0}\right) -\left[ \mu \left(
A^{2}\Gamma _{m,n,1,1,0,0}\right) \right] ^{2}\},  \tag{B8}
\end{align}
and
\begin{align}
C_{Q_{12}}& =4[\mu ^{2}\left( A^{2}\Gamma _{m,n,1,1,1,1}\right)  \notag \\
& -\mu ^{2}\left( A^{2}\Gamma _{m,n,1,1,0,0}\right) \left( A^{2}\Gamma
_{m,n,0,0,1,1}\right) ],  \tag{B9}
\end{align}
\begin{equation}
C_{Q_{21}}=C_{Q_{12}},  \tag{B10}
\end{equation}%
as well as
\begin{align}
C_{Q_{22}}& =4\{ \mu ^{2}\left( A^{2}\Gamma _{m,n,0,0,2,2}+A^{2}\Gamma
_{m,n,0,0,1,1}\right)  \notag \\
& +\gamma \left( A^{2}\Gamma _{m,n,0,0,1,1}\right) -\left[ \mu \left(
A^{2}\Gamma _{m,n,0,0,1,1}\right) \right] ^{2}\}.  \tag{B11}
\end{align}

The calculation results of intramodal and intermodal correlations are as
follows%
\begin{equation}
g_{a}^{\left( 2\right) }=\frac{A^{2}\Gamma _{m,n,2,2,0,0}}{\left(
A^{2}\Gamma _{m,n,1,1,0,0}\right) ^{2}},  \tag{B12}
\end{equation}%
and
\begin{equation}
g_{b}^{\left( 2\right) }=\frac{A^{2}\Gamma _{m,n,0,0,2,2}}{\left(
A^{2}\Gamma _{m,n,0,0,1,1}\right) ^{2}},  \tag{B13}
\end{equation}%
as well as%
\begin{equation}
g_{ab}^{\left( 2\right) }=\frac{A^{2}\Gamma _{m,n,1,1,1,1}}{\left(
A^{2}\Gamma _{m,n,1,1,0,0}\right) \left( A^{2}\Gamma _{m,n,0,0,1,1}\right) }.
\tag{B14}
\end{equation}

\end{document}